\documentclass[longauth]{aa} 

\usepackage{graphicx}
\usepackage{txfonts}
\usepackage{breqn}
\newcommand{\MJup}{$M_\mathrm{Jup}$}

\newcommand{\logg}{$\log g$}

\newcommand{\Ks}{$K_\mathrm{s}$}
\newcommand{\Lp}{$L^\prime$}

\begin{document}

   \title{Isochronal age-mass discrepancy of young stars: SCExAO/CHARIS integral field spectroscopy of the HIP 79124 triple system}


\author{Ruben Asensio-Torres\inst{1}, 
        Thayne Currie\inst{2,3,4}, 
Markus Janson\inst{1},
Silvano Desidera\inst{5},
Masayuki Kuzuhara\inst{6, 7},
Klaus Hodapp\inst{8},
Timothy D. Brandt\inst{9},
Olivier Guyon\inst{3,6,10,11},
Julien Lozi\inst{3},
Tyler Groff\inst{12},
Jeremy Kasdin\inst{13},
Jeffrey Chilcote\inst{14},
Nemanja Jovanovic\inst{15},
Frantz Martinache\inst{16},
Michael Sitko\inst{17},
Eugene Serabyn\inst{18},
Kevin Wagner\inst{10},
Eiji Akiyama\inst{19},
Jungmi Kwon\inst{20},
Taichi Uyama\inst{21},
Yi Yang\inst{22},
Takao Nakagawa\inst{20},
Masahiko Hayashi\inst{7},
Michael McElwain\inst{12},
Tomoyuki Kudo\inst{3},
Thomas Henning\inst{23},
Motohide Tamura\inst{6,7,24}
} 

\authorrunning{R. Asensio-Torres et al.}
\titlerunning{HIP 79124}

\institute{Department of Astronomy, Stockholm University, AlbaNova University Center, SE-106 91 Stockholm, Sweden
  \and NASA-Ames Research Center, Moffett Field, California 94035 
  \and National Astronomical Observatory of Japan, Subaru Telescope, National Institutes of Natural Sciences, Hilo, HI 96720, USA 
  \and Eureka Scientific, 2452 Delmer Street Suite 100. Oakland, CA 94602-3017 
  \and INAF-Osservatorio Astronomico di Padova, Vicolo dell’Osservatorio 5, 35122 Padova, Italy 
  \and Astrobiology Center, National Institutes of Natural Sciences, 2-21-1 Osawa, Mitaka, Tokyo, Japan 
  \and National Astronomical Observatory of Japan, 2-21-1, Osawa, Mitaka, Tokyo, 181-8588, Japan
  \and Institute for Astronomy, University of Hawaii, Hilo, HI 
  \and Department of Physics, University of California-Santa Barbara, Santa Barbara, CA 
  \and Steward Observatory, University of Arizona, Tucson, AZ 85721, USA 
  \and College of Optical Sciences, University of Arizona, Tucson, AZ 85721, USA 
  \and NASA-Goddard Space Flight Center, Greenbelt, MD, USA 
  \and Department of Mechanical Engineering, Princeton University, Princeton, NJ, USA
  \and Department of Physics, University of Notre Dame, 225 Nieuwland Science Hall, Notre Dame, IN, 46556, USA 
  \and Department of Astronomy, California Institute of Technology, 1200 E. California Blvd., Pasadena, CA 91125, USA
  \and Universite Cote d’Azur, Observatoire de la Cote d’Azur, CNRS, Laboratoire Lagrange, France
  \and Center for Extrasolar Planetary Systems, Space Science Institute, 1120 Paxton Ave., Cincinnati, OH 45208, USA
  \and Jet Propulsion Laboratory, California Institute of Technology, 4800 Oak Grove Drive, Pasadena, CA 91109, USA
  \and Institute for the Advancement of Higher Education, Hokkaido University, Kita 17, Nishi 8, Kita-ku, Sapporo, 060-0817, Japan
  \and ISAS/JAXA, 3-1-1 Yoshinodai, Chuo-ku, Sagamihara, Kanagawa 252-5210, Japan
  \and Department of Astronomy, Graduate School of Science, The University of Tokyo, 7-3-1, Hongo, Bunkyo-ku, Tokyo, 113-0033, Japan
  \and Department of Astronomy, The Graduate University for Advanced Studies, National Astronomical Observatory of Japan
  \and Max Planck Institut f\"{u}r Astronomie, K\"{o}nigstuhl 17, 69117 Heidelberg, Germany
  \and  Department of Astronomy, Graduate School of Science, The University of Tokyo, 7-3-1, Hongo, Bunkyo-ku, Tokyo, 113-0033, Japan
}


 \abstract{ We present SCExAO/CHARIS 1.1--2.4 $\mu$m integral field direct spectroscopy of the young HIP 79124 triple system. HIP 79124 is a member of the Scorpius-Centaurus association, consisting of an A0V primary with two low-mass companions at a projected separation of $<$1\,$\arcsec$. Thanks to the high quality wavefront corrections provided by SCExAO, both companions are decisively detected without the employment of any PSF-subtraction algorithm to eliminate quasi-static noise. The spectrum of the outer C object is very well matched by Upper Scorpius M4\,$\pm$\,0.5 standard spectra, with a T$\rm_{eff}$ = 2945\,$\pm$\,100\,K and a mass of $\sim$\,350\,M$\rm_{Jup}$. HIP 79124 B is detected at a separation of only 180\,mas in a highly-correlated noise regime, and it falls in the spectral range M6\,$\pm$\,0.5 with T$\rm_{eff}$ = 2840\,$\pm$\,190\,K and $\sim$\,100\,M$\rm_{Jup}$. Previous studies of stellar populations in Sco-Cen have highlighted a discrepancy in isochronal ages between the lower-mass and higher-mass populations. This could be explained either by an age spread in the region, or by conventional isochronal models failing to reproduce the evolution of low-mass stars. The HIP 79124 system should be coeval, and therefore it provides an ideal laboratory to test these scenarios. We place the three components in a color-magnitude diagram and find that the models predict a younger age for the two low-mass companions ($\sim$\,3\,Myr) than for the  primary star ($\sim$\,6\,Myr). These results imply that the omission of magnetic effects in conventional isochronal models inhibit them from reproducing early low-mass stellar evolution, which is further supported by the fact that new models that include such effects provide more consistent ages in the HIP 79124 system.
  } 


\maketitle 

%

\section{Introduction}
A large number of direct imaging surveys searching for substellar companions have been performed in recent years, yielding the first directly-imaged exoplanets \citep{bowler2016}. The efforts have been put into looking for these objects around young and nearby stars, where contrast ratios are more favorable. These discoveries reveal colors and spectral features that can indicate the composition of their atmospheres and their underlying physical properties \citep[e.g.,][]{barman2011,currie2011,faherty2016, biller2018}. \par

The Scorpius Centaurus (Sco-Cen) region is the nearest OB association \citep[$\sim$\,100--200\,pc,][]{dezeeuw1999}, and contains young stars ranging from hot and very massive O-type stars to free-floating substellar objects \citep{cook2017}. It is thus an ideal place in which to search for young and low-mass objects,  allowing a detailed study of stellar evolution and planet formation mechanisms \citep[e.g.][]{preibisch2002,preibisch2008, luhman2012,currie2015,pecaut2016}.\par

The large number of stars present in this association has been used to statistically constrain the age of each subregion by comparing stars of different masses to theoretical isochrones, which has shown an intricate star formation history \citep[e.g.,][]{pecaut2016}. One of the most alluring results is the existence of a mass-dependent age trend, i.e., models yield younger ages for cooler low-mass stars compared to the corresponding massive population in the same subregion. For instance, the Upper Sco (USco) subregion would have a median age of about 5\,Myr if we consider only the pre-main sequence (PMS) K- and early M-type population, which is half of the usually adopted USco age of 10\,$\pm$3\,Myr from PMS G- and F-type stars. This discrepancy has also been similarly observed in other regions \citep[e.g.,][]{hillebrand1997, bell2015, herczeg2015}. Two main explanations have been proposed to explain this fact; the effect of magnetic fields, which might inhibiting convection and slow down the contraction rate in low-mass PMS stars \citep{feiden2016, somers2017}, or an age spread within young clusters \citep{fang2017}.\par

Here we present a SCExAO/CHARIS spectroscopic study of the USco HIP 79124 system encompassing the $JH$\Ks\, near-IR bands (1.1--2.4\,$\mu$m). This young A-type star forms a triple system with two resolved low-mass companions at a very close projected separation. The outer companion, HIP 79124 C, was discovered at about 1\,\arcsec ($\sim$\,137\,AU) by AO-assisted direct imaging surveys, first by \citet{kou2005,kou2007} with ADONIS, which flagged it as a companion, and later by \citet{lafre2014} with NIRI at the Gemini North telescope. Recently, an even more interesting discovery was the presence of an additional $\sim$135\,\MJup\, companion, HIP 79124 B, interior to the C component, using aperture masking interferometry in \Lp\,band \citep{hink2015}. This companion has also been imaged for the first time at a separation of only 0.18\,\arcsec ($\sim$\,25\,AU) by \citet{ser2017}, using the new \Lp\,optical vortex coronagraph working alongside the AO-assisted NIRC2/Keck camera.\par

In this work we will make use of our new low-resolution SCExAO/CHARIS observations together with archival Keck/NIRC2 \Lp\,photometry to analyze, for the first time, the spectral nature of the B and C companions. This study will improve the knowledge on the parameters of two low-mass stars in USco. Moreover, star-formation models favors a scenario where the HIP 79124 triple system is coeval, as the time difference of massive A-type formation compared to low-mass M-type objects seems to be securely below 1\,Myr, even after including the accretion phase \citep{bate2012}. Binary systems in the Taurus association have also been shown to be more coeval than the region as a whole \citep{kraus2009}. The likely coevality makes of this system a perfect laboratory to test whether the conventional isochronal models predict the same age for both the two low-mass companions and the massive and hot primary A-type star, or otherwise fail at reproducing the PMS evolution of low-mass stars.

\begin{table*}
\caption{\label{tab:obslog}HIP 79124 Observing Log.}
\small
\centering
\begin{tabular}{lccccccc}
\hline\hline
        Date  
       &Telescope/Instrument    
       &  Wavelength 
       & Coronagraph 
       &  $t\rm_{int}$
       & $N\rm_{images}$   
       &Collected PA 
       & Mode
       \\
       (UT)
       &
       & (Filter)
       &
       & (sec)
       &
       & (deg)
       &\\
\hline
2017-07-15 & Subaru/CHARIS & 1.1--2.4 $\mu m$ & Lyot & 31\,$\times$1.475 & 15  & 4.3 & IFS\\
\\
2017-07-15 & Subaru/CHARIS & 1.1--2.4\,$\mu$m + ND filter  & -- & 100\,$\times$0.2 & 3 & 2.0 & IFS\\

\\
\textit{Archival Data}\\
2016-04-13 & Keck/NIRC2 & \Lp (3.8\,$\mu$m) & Vortex & 30 & 26 & 18.8  & Imaging \\
\hline
\end{tabular}
\tablefoot{The archival 2016 Keck/NIRC2 data was published in \citet{ser2017}.\\
\tablefoottext{a}{Photometry for MF13, LS~1267 and HD~80077 from
Dupont et al.}
\tablefoottext{b}{Photometry for LS~1262, LS~1269 from
Durand et al.}
\tablefoottext{c}{Photometry for MO2-119 from
Mathieu et al.}
}
\end{table*}

\section{Observations and data reduction}
\label{sec:observations}

\subsection{SCExAO/CHARIS Data} 
We observed the HIP 79124 system with the newly-established Coronagraphic High Angular Resolution Imaging Spectrograph (CHARIS) on 2017-07-15, located on the Nasmyth platform at the Subaru Telescope in Hawaii \citep{peters2012,groff2013} coupled to the extreme adaptive optics system SCExAO \citep{jovanovic2015}. We used the low-resolution (R $\sim$\,20) configuration of CHARIS, which covers the $J$ + $H$ + $K$ bands within a field of view (FoV) of 2.2\,$\arcsec$ $\times$ 2.2\,$\arcsec$, and collected all data in pupil tracking/angular differential imaging mode \citep[ADI,][]{marois2006}.  Both seeing conditions (e.g., $\theta$ $\sim$ 0.8\,$\arcsec$) and AO performance  were slightly below average but sufficient to reveal both HIP 79124 companions without post-processing.

   \begin{figure}
   \centering
   \includegraphics[width=\hsize]{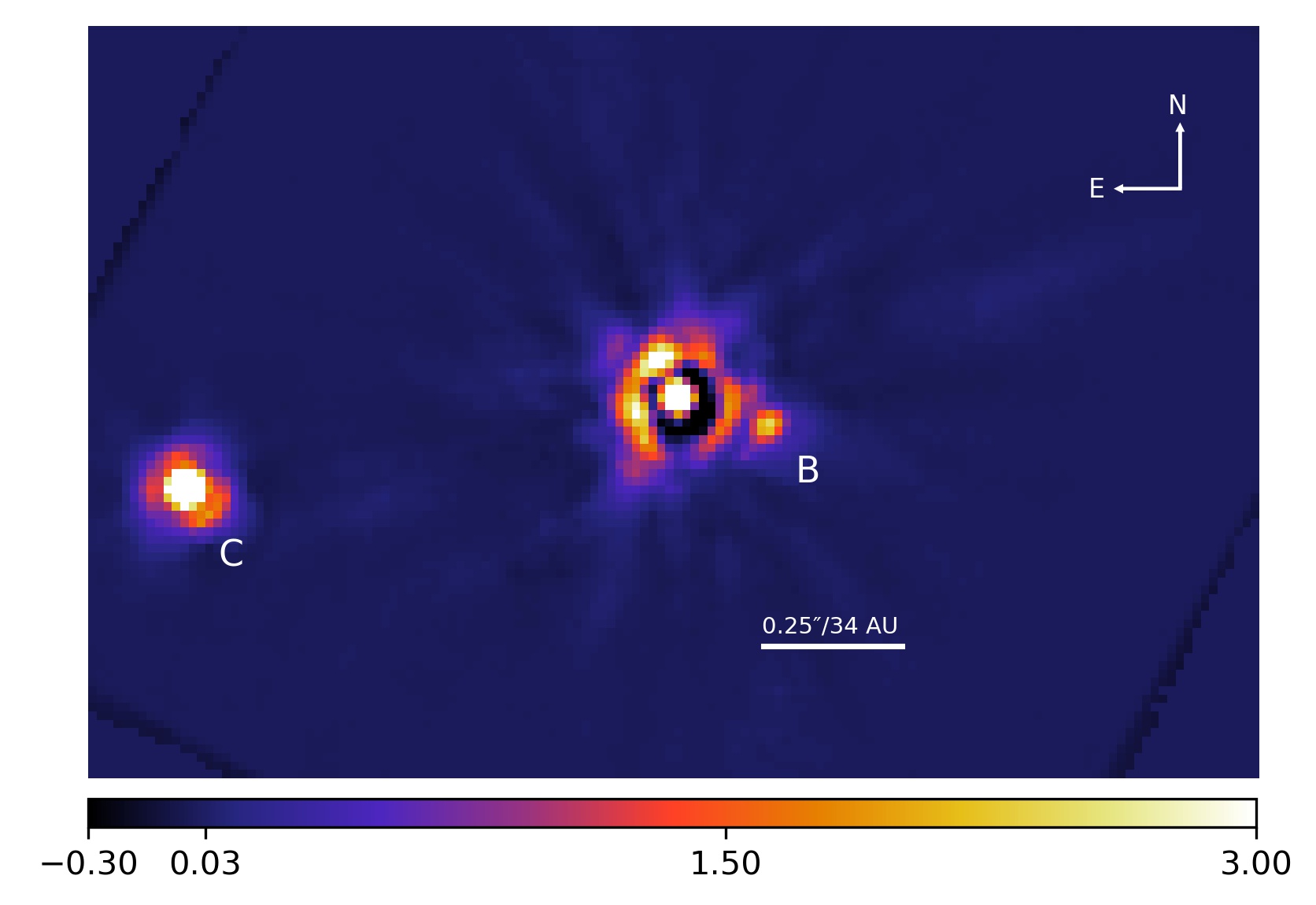}
      \caption{Wavelength-collapsed $JH$\Ks CHARIS image of the HIP 79124 stellar system. After subtracting a radial profile, both the B and C low-mass companions are clearly detected at a S/N of $\sim$9 and $\sim$120, respectively. Another set of non-coronagraphic data was used to extract the spectrum of the C companion. 
              }
         \label{fig:HIP_79124}
   \end{figure}

The observations consisted of two sequences.  First, we acquired a first set of images using the Lyot coronagraph with a 217\,mas diameter occulting spot to block the star, each of which consisting of 31 coadded 1.475\,sec exposures. After frame selection, our total integration time was $\sim$\,11 minutes, covering a parallactic angle motion of 4.3\,$\deg$. As HIP 79124 C was easily visible but possibly in the non-linear regime of the detector in a few spectral channels in the coronagraphic data, we acquired a second set of non-coronagraphic data using a neutral density filter to keep both A and C in the linear count regime totalling 60\,sec of integration time and covering a smaller parallactic angle motion of $\sim$\,2\,$\deg$.
Table \ref{tab:obslog} shows the observation log of the CHARIS data and the archival NIRC2/Keck data that we retrieve to complement our analysis (see Section \ref{sec:keck}).\par 

For data cube extraction, we used the 'least squares' method presented in the CHARIS Data Reduction Pipeline \citep[CHARIS DRP;][]{brandt2017}, yielding 22 images at wavelengths between 1.1--2.4\,$\mu$m. An IDL-based CHARIS data reduction pipeline was used to perform the basic reduction processes, such as background subtraction, flat fielding and image registration \citep{currie2018b} following previous methods applied for broadband imaging data \citep[e.g.,][]{currie2011}. To spectrophotometrically calibrate each data cube, we used an A0V spectral type from the \citet{pickles1998} stellar library, which is shown to be accurate despite issues with the library at other spectral types \citep{currie2018b} and is also adopted in the GPI Data Reduction Pipeline \citep{perrin2014}.  For the unsaturated data, the star's photometry is used directly for flux-calibration. For the coronagraphic data, satellite spots provided absolute spectrophotometic calibration \citep[e.g.,][]{currie2018a}. We verified and fine-tuned our  spectrophotometric calibration by 1) comparing the brightness of HIP 79124 C in the saturated and unsaturated data and 2) correcting for signal loss due to the Lyot occulting spot ($\sim$ 92.5\,\% throughput at $\rho$ $\sim$\,0.18\,$\arcsec$).\par

\subsection{Keck/NIRC2 $L_{\rm p}$ Archival Data}
\label{sec:keck}
The \Lp magnitude of the B component was already presented in \citet{hink2015} and \citet{ser2017}; however, neither paper reported photometry for HIP 79124 C. To obtain it, we reduced the archival Keck/NIRC2 \Lp vortex data from June 2015 and April 2016 published in \citet{ser2017}, focusing on acquisition frames where the primary and C are both unsaturated. Basic processing followed previous steps employed for reducing NIRC2 \Lp archival data \citep{currie2011,currie2018b}, including a linearity correction, sky subtraction and distortion correction. The results from both epochs were consistent within error bars, and we adopted the smaller photometric uncertainty.

\subsection{SCExAO/CHARIS Detections}

In Figure \ref{fig:HIP_79124} we show the wavelength-collapsed image of HIP 79124 obtained from combining our coronagraphic exposures. Given that little field rotation was collected, we refrain from applying a PSF-subtraction algorithm that benefits from the ADI observing technique. Instead, we simply subtract a radial profile, which is sufficient to recover the B companion at a decent signal-to-noise ratio (S/N).\par

To compute the S/N for both the B and C objects, we divide the convolved flux, measured at each companion's position, by the standard deviation of the convolved residual noise in concentric annuli at their same separation after excluding the signal from the companion \citep[e.g.,][]{thalmann2009,currie2011}, and corrected for finite sample sizes \citep{Mawet2014}. Our simple wavelength-collapsed radial profile-subtracted image yields S/N\,$\sim$\,9 and $\sim$\,120, respectively for B and C. As expected, the brightness and separation of the C companion provides a very strong detection in all individual channels. Although the situation is more complicated for the very close B object, SCExAO/CHARIS is able to identify it with a S/N of $\sim$\,6--8 for the shortest wavelengths in the data cube, peaking at 2--2.2\,$\mu$m with a S/N of $\sim$\,11. \par

From Figure \ref{fig:HIP_79124} we also obtain the astrometric position of the unsaturated companions with respect to the primary. We perform a Gaussian fit to extract the centroid of the point sources, whose error is estimated by dividing the FWHM of the candidate's PSF by its S/N \citep{thalmann2014}. To this uncertainty, we add those from the plate scale of 16.2\,$\pm$\,0.1\,mas/spaxel and true north orientation of -2.20\,$\pm$0.27\,$\deg$ \citep{currie2018b}. The inner companion is detected at a projected separation of  $\rho$ = 180\,$\pm$\,5\,mas ($\sim$\,25\,AU) and a position angle of 252.9\,$\pm$\,1.6\,$\deg$. Within error bars, the angular separation is consistent with those obtained by \citet{hink2015} in April 2010 and \citet{ser2017} in April 2016. The position angle increases in time over these previous measurements, supporting evidence for counterclockwise orbital motion first found by \citet{ser2017}. We find C at 0.967$\pm$\,0.006\,\arcsec ($\sim$\,132\,AU) and 100.39$\pm$\,0.04\,$\deg$. Using the archival astrometric data from \citet{lafre2014}, we confirm that the C companion shares common proper motion with the primary star.




\section{Extracted CHARIS spectra and photometry}

\begin{figure}
\includegraphics[width=\hsize]{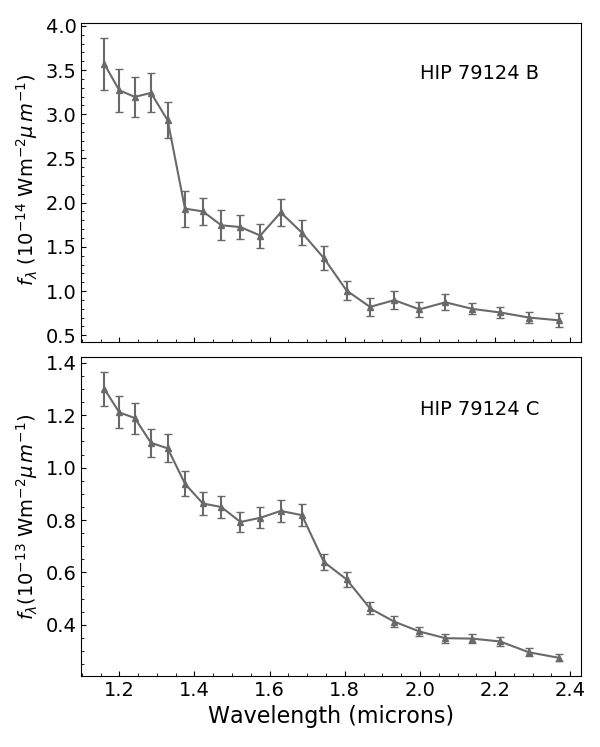}
\caption{SCExAO/CHARIS $JH$\Ks \,spectra of the HIP 79124 B and C companions. The flux has been calibrated with an A0V template from \citet{pickles1998} and dereddened by $A_{V}$ = 0.82.  }
\label{fig:spectra}
\end{figure}

\subsection{Spectrophotometry and reddening}
\label{sec:specandphot}

Figure \ref{fig:spectra} shows the CHARIS low-resolution extracted spectra for the HIP 79124 B and C  companions in units of \ Wm$^{-2}$$\mu$m$^{-1}$.\ We carried out the extraction by defining an aperture of 0.5\,$\lambda$/D around the position of each companion in the wavelength-collapsed image.

The B and C spectra present very similar features. They show a downward trend in flux toward redder wavelengths, except for a peak or plateau in the $H$ band region ($\lambda$\,$\sim$\,1.65\,$\mu$m). The C companion is brighter and detected with a very high S/N, which generate very small error bars during the extraction.
In this case, the error bars are dominated by the absolute calibration uncertainty. \par

The photometry of the triple system is presented in Tables 2 and 3. We convolved each spectrum with the Mauna Kea Observatory (MKO) $JHK_{s}$ passbands functions, previously binned down to CHARIS' low-resolution mode. We note that the photometric values for the C companion are very close to the ones given by \citet{kou2007}, although the $J$ band photometry is slightly brighter.\par


The $JHK\rm_{s}$ 2MASS magnitudes of the primary were converted to the MKO system by means of the transformation equations in \citet{carpenter2001}. We checked that the C companion does not affect the 2MASS photometry significantly, as the difference between including or not including the flux of the companion in the magnitude
of the primary is within 2\,$\sigma$ of the primary error bar in the K$\rm_{s}$ and \Lp bands.\par

We also estimated the extinction for our target using the intrinsic color of nearby dwarf A0-type stars with negligible extinction \citep{pecaut2013}. We computed $E(B-V)$, $E(V-J)$, $E(V-H)$ and $E(V-K$\rm{s}$)$. 
Taking R$\rm_{V}$ = 3.1 as extinction law, and the extinction coefficients from \citet{fiorucci2003}, we obtain a median $A_{V} = 0.82 \pm 0.05$ mag, $A_{J} = 0.23 \pm 0.02$ mag, $A_{H} = 0.136 \pm 0.008$ mag and $A_{Ks} = 0.088 \pm 0.005$ mag from the four different colors and adopt the scatter as uncertainties. This  $A_{V}$ value is in agreement within errorbars with the values previously obtained by \citet{pecaut2012} and \citet{hink2015}.  No extinction is assumed in \Lp, and we use these values to derive dereddened absolute magnitudes for a GAIA-DR2 distance of 137.0\,$\pm$\,1.2\,pc \citep{lindegren2018}. We dereddened the spectra of the B and C companions by fitting a second-order polynomial to the extinction coefficients, obtaining a coefficient for each of the CHARIS wavelength channels, which we used to correct for the reddening in our spectra.\par

\begin{table*}[!htbp]
\caption{Photometry of HIP 79124 B\label{tab:photometryB} }
\centering
\small
\begin{tabular}{l l c c c c c l }
\hline\hline
UT Date           & Telescope/Camera          & Filter            & Primary                  &Companion                  & Apparent mag     & Absolute mag & Ref.  \\
 	            	&	           	 &               & (mag)          &($\Delta$mag)            &   (mag)         & (dereddened)       &        \\
\hline                      
2016-04-13 & Keck/NIRC2 & W1/\Lp       & 6.96\,$\pm$\,0.04\tablefootmark{a} &   4.25\,$\pm$\,0.14\tablefootmark{c}       & 11.16\,$\pm$\,0.11 & 5.48\,$\pm$\,0.11 &  2,3  \\  

2017-07-15 & SCExAO/CHARIS & $J$       & 7.17\,$\pm$\,0.03\tablefootmark{b} & 5.48\,$\pm$\,0.13        & 12.65\,$\pm$\,0.13 & 6.73\,$\pm$\,0.13  &  1  \\  

2017-07-15 & SCExAO/CHARIS & $H$       & 7.00\,$\pm$\,0.05\tablefootmark{b} & 5.26\,$\pm$\,0.15       & 12.26\,$\pm$\,0.15 & 6.44\,$\pm$\,0.15  &  1  \\ 

2017-07-15 & SCExAO/CHARIS & $Ks$      & 7.003\,$\pm$\,0.018\tablefootmark{b} & 4.92\,$\pm$\,0.15       & 11.93\,$\pm$\,0.15 & 6.15\,$\pm$\,0.15 &  1  \\

\hline
\end{tabular}
\tablebib{(1)~This paper;
(2)  \citealt{ser2017}; (3) \citealt{hink2015}.
}
\tablefoot{\tablefoottext{a}{From the WISE W1 channel \citep{cutri2012}}
\tablefoottext{b}{From the 2MASS catalog \citep{cutri2003} converted to the MKO passbands via \citet{carpenter2001}}
\tablefoottext{c}{Mean of the magnitude contrast published in \citet{hink2015} and \citet{ser2017}.}

}
\end{table*}

\begin{table*}[!htbp]
\caption{Photometry of HIP 79124 C\label{tab:photometryC} }
\centering
\small
\begin{tabular}{l l c c c c c l }
\hline\hline
UT Date           & Telescope/Camera          & Filter            & Primary                  &Companion                  & Apparent mag     & Absolute mag & Ref.  \\
 	            	&	           	 &               & (mag)          &($\Delta$mag)            &   (mag)         & (dereddened)       &        \\
\hline                      
2016-04-13 & Keck/NIRC2 & W1/\Lp  & 6.96\,$\pm$\,0.04\tablefootmark{a} & 2.98\,$\pm$\,0.03      & 9.94\,$\pm$\,0.05 & 4.26\,$\pm$\,0.05 &  1,2  \\  

2017-07-15 & SCExAO/CHARIS & $J$       & 7.17\,$\pm$\,0.03\tablefootmark{b} & 4.08\,$\pm$\,0.01         & 11.27\,$\pm$\,0.05 & 5.36\,$\pm$\,0.05 & 1  \\  

2017-07-15 & SCExAO/CHARIS & $H$       & 7.00\,$\pm$\,0.05\tablefootmark{b} & 3.54\,$\pm$\,0.01          & 10.57\,$\pm$\,0.05 &  4.75\,$\pm$\,0.05 &  1 \\

2017-07-15 & SCExAO/CHARIS & $Ks$       & 7.003\,$\pm$\,0.018\tablefootmark{b} & 3.344\,$\pm$\,0.01         & 10.35\,$\pm$\,0.03 & 4.58\,$\pm$\,0.05 &  1 \\

\hline
\end{tabular}
\tablebib{(1)~This paper;
(2)  \citealt{ser2017}.}
\tablefoot{\tablefoottext{a}{From the WISE W1 channel \citep{cutri2012}};
 \tablefoottext{b}{From the 2MASS catalog \citep{cutri2003} converted to the MKO passbands via \citet{carpenter2001}.}

}
\end{table*}

\subsection{Correlated noise}

When fitting integral field spectrograph (IFS) data to model spectra, it is of vital importance to consider the effect of covariances to properly retrieve atmospheric parameters. As shown by \citet{greco2016}, in high-contrast imaging it is not easy to understand the impact that the data analysis techniques have on the extracted spectrum. Following their work, we first compute the spectral correlation at HIP 79124 B's and C's positions in the collapsed image. To this aim, we normalize each channel by its standard deviation profile and, for each pair of CHARIS wavelengths $i$ and $j$, compute its correlation $\Psi\rm_{ij}$ in a 2\,$\lambda$/D-wide ring centered on the companion:

\begin{equation}
\Psi_{ij} \equiv  \frac{<I_{i}\,I_{j}>}{\sqrt{<I_{i}^{2}><I_{j}^{2}>}}
\label{eq:phi_corr}    
\end{equation}

where $I\rm_{i}$ and  $I\rm_{j}$ are the intensities at wavelengths $i$ and $j$, respectively, averaged over all the spatial locations within the annulus, and masking the 2\,$\lambda$/D region around the companion.\par

\begin{figure}
\setlength{\unitlength}{\textwidth}
\hspace*{-0.45cm}                                                           
 \includegraphics[width=0.56\textwidth]{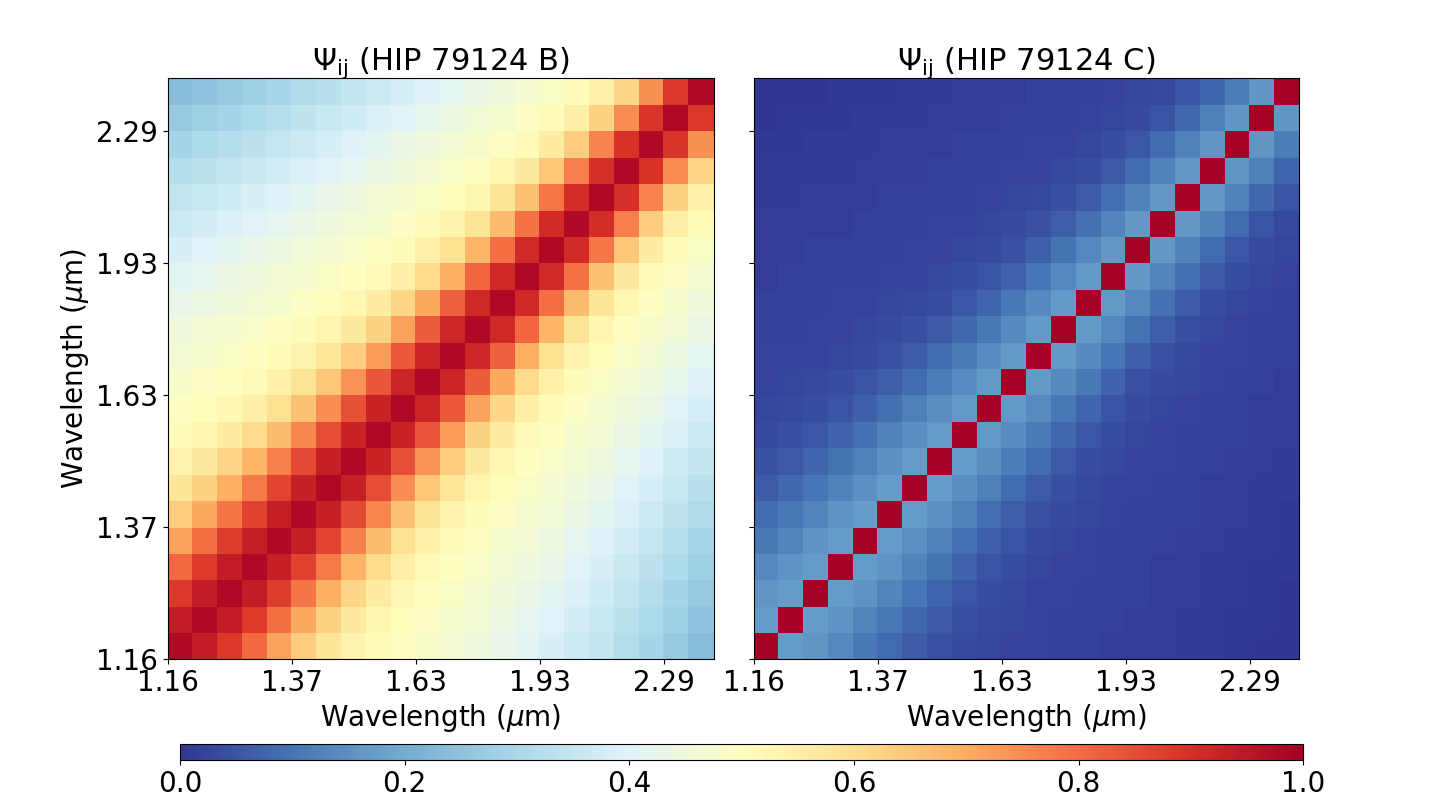}

\caption{Correlation matrices calculated via Eq. \ref{eq:phi_corr} for the B (left) and C (right) companions to HIP 79124 A. The reduction process of the IFS data introduced spectrally correlated noise in the extracted spectra. The small projected separation at which B is located makes the noise to be highly-correlated, as the speckles from the primary PSF add flux density within the companion's PSF over different channels.}
\label{fig:corrmatrices}
\end{figure}

The correlation matrices for both companions are shown in Figure \ref{fig:corrmatrices}. The noise is very much uncorrelated at the position of HIP 79124 C ($\sim$\,1\,$\arcsec$), with a nearly constant value of $\sim$\,0.2 except for $i=j$. This might be due to the fact that we used the unsaturated raw image for spectral extraction of this outer companion. For the inner companion at $\sim$\,0.18\,$\arcsec$, HIP 79124 B, the noise is highly correlated and the correlation varies between channels. \par

The populated $\Psi\rm_{ij}$ vs $\lambda\rm_{ij}$ distribution can be fit using a Levenberg-Marquardt minimization to the functional form described in \citet{greco2016}:

\begin{dmath}
\Psi_{ij} \approx A_{\rho}\,exp\left[-\frac{1}{2}\left(\frac{\rho}{\sigma_{\rho}}\frac{\lambda_{i} - \lambda_{j}}{\lambda_{c}} \right)^{2}\right] + \\
A_{\lambda}exp\left[ -\frac{1}{2}\left(\frac{1}{\sigma_{\lambda}}       \frac{\lambda_{i} - \lambda_{j}}{\lambda_{c}}\right)^2 \right] + \\
A_{\delta}\delta_{ij} 
\end{dmath}

where the spatially and spectrally correlated noise have amplitudes A$_{\rho}$ and A$_{\lambda}$, and characteristic lengths $\sigma_{\rho}$ and $\sigma_{\lambda}$, respectively. A$_{\delta}$ is the amplitude of the uncorrelated term.\par
We redo this minimization several times at separations bracketing the companions, masking them and any residual satellite spot light. In the case of the B companion, it appears that the spectral covariance is not optimally described by this functional form, as there are two secondary peaks in $\Psi_{ij}$ that flank the main peak at $i=j$. Comparing plots for different separations, these peaks move out from $\rho(\lambda_{i} - \lambda_{j})/(\lambda_{c})$ $\sim$\,1.5 to 2.5, as we go from separations of $\rho$ = 4 to 6 in units of $\lambda\rm_{c}$/D. The peaks disappear at $\rho$ = 3 and $\rho$\,$>$\,7. In any case, our best fit for the B companion at its  separation of $\rho$ = 4.25, is A$_\rho$ = 0.42, A$_\lambda$ = 0.54, A$_\delta$ = 0.02, $\sigma_\rho$ = 0.34 and $\sigma_\lambda$ = 0.57. This shows that at the location of the B object, the uncorrelated component A$_\delta$ is very much negligible compared to the correlated noise, which is dominant at small separations where speckle noise is not well eliminated. In the case of the C companion the uncorrelated amplitude dominates at $\rho$ = 4.25, with A$_\rho$ = 0.12, A$_\lambda$ = 0.21, A$_\delta$ = 0.65, $\sigma_\rho$ = 2.15 and $\sigma_\lambda$ = 1.69. As we extracted the C spectrum from the unsaturated dataset, the resulting correlation length is large perhaps because the background has not been well flattened from a least-squares PSF subtraction.

\section{HIP 79124 empirical constraints}\label{sec:models}

\subsection{Spectral Types of HIP 79124 B and C}

Although we have adopted an A0V spectral type for the primary star from \citet{houk1988}, the spectral type of the companions has never been assessed. To get a good estimation, we first compare the CHARIS near-IR spectrum of HIP 79124 B and C to libraries of ML empirical objects in young moving groups. Then, we adopt a second approach, refining the classification by comparing the observed HIP 79124 spectra to a set of $\sim$\,10\,Myr-old M-type standard spectra.

\subsubsection{Comparison to empirical ML spectra}
\label{sec:empirical}
As HIP 79124 is a member of the USco young star-forming region with an age of only 10\,$\pm$\,3\,Myr \citep{pecaut2016}, we decide to use libraries of young objects, which are warmer than their field counterpart (and have earlier spectral type) for a given mass, and are still contracting, presenting inflated radii and thus low surface gravities that affect their spectra. An indication of youth in low-resolution near-IR spectra is the triangular $H$-band continuum shape, which becomes less pronounced as one moves from very low ($\delta$) to low ($\gamma$) and intermediate-gravity ($\beta$) late $M$- and $L$-type dwarfs. In comparison, field objects tend to show a plateau \citep[e.g.,][]{allers2013}. 
Other indicators exist also in the $J$ and $K$ bands, such as FeH absorption \citep{mclean2003} or the $K$-band slope \citep[see the H$\rm_{2}$($K$) index,][]{canty2013,currie2014a}.\par

We mainly adopt the young population of the Montreal Spectral library \footnote{\url{https://jgagneastro.wordpress.com/the-montreal-spectral-library/}} as the source for comparison spectra. These objects are members of nearby young moving groups ($\leq$\,120\,Myr), with spectral types in the $MLT$ range and $\delta$, $\gamma$ and $\beta$ gravities.  We consider only high S/N objects, leaving out those with median uncertainties larger than 5\,$\%$ of the median flux value. 
These spectra come mainly from \citet{gagne2015} and were obtained with several instruments, such as $Flamingos-2$ \citep{eikenberry2004} and SpeX \citep{rayner2003}. Also, we include the near-IR \citet{bonnefoy2014} VLT/SINFONI library of young dwarfs in the $M-L$ transition (M8.5--L4). \par

In the chi-square goodness of fit statistic we incorporate the correlated errors via the covariance matrix $C$:

\begin{equation}
\chi^{2}_{k} = (S - f_{k}F_{k})^{T} C^{-1}_{k} (S - f_{k}F_{k})
\end{equation}

where $S$ is the set of observed spectral values by CHARIS. To create a similar set of these flux values for each of the comparison spectra, we first smooth the empirical model to the CHARIS low resolution. Then, for each of the 22 CHARIS channels spanning the 1.1 -- 2.4\,$\mu$m range, we estimate a flux value via interpolation. In this way we create a vector $F$ for each empirical $k$ object that will be compared to $S$. The comparison object is multiplied by a constant $f\rm_{k}$, which was introduced in \citet{cushing2008}, that minimizes $\chi^{2}$ and accounts for the distance difference between the observed CHARIS spectrum and the empirical object. We focus on regions covering the major $JHK$ passbands, avoiding strong telluric absorption, leaving us with 15 out of the 22 channels that will be used for the comparison.\par

To construct the 22\,$\times$\,22 covariance matrix $C$, we use the correlation matrix $\Psi\rm_{ij}$ presented in Figure \ref{fig:corrmatrices} for the B and C companions. The off-diagonal elements of $\Psi\rm_{ij}$ are multiplied by the uncertainty in the flux of the corresponding $ij$ channels. For the outer companion, the error bars from the extracted spectrum are on the order of 0.1\,$\%$, and the absolute calibration uncertainty dominates, with values 2--4\,$\%$ of the observed flux. The on-diagonal are also affected by the uncertainties on the uncorrelated comparison spectrum, which we add in quadrature to the correlated errors.\par

\begin{figure}
\setlength{\unitlength}{\textwidth}
\hspace*{-0.3cm}                                                                                                  
 \includegraphics[width=0.525\textwidth]{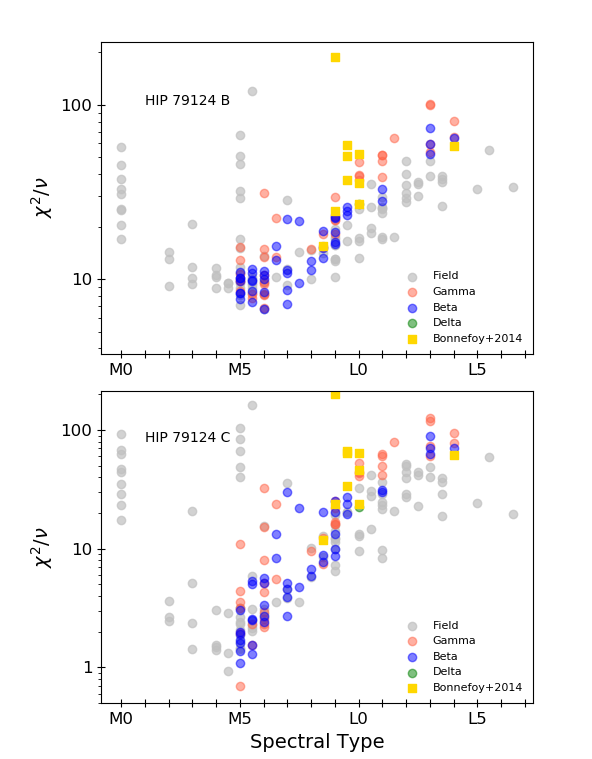}

\caption{Normalized $\chi^{2}$ for the B and C CHARIS spectra as compared to the empirical objects of the Montreal \citep{gagne2015} and Bonnefoy-VLT/SINFONI \citep{bonnefoy2014} libraries.}
\label{fig:chisq_young}
\end{figure}


Figure \ref{fig:chisq_young} shows the resulting $\chi^{2}$ per degree of freedom for the Montreal and Bonnefoy libraries, filtered out for objects with a signal to noise lower than 5\,\%. For the outer C companion, M3--M5.5 objects fall within the $\Delta$\,$\chi^{2}$ 95\,$\%$ confidence level for 15 degrees of freedom. Low-gravity comparison objects with spectral types earlier than M5 are unavailable.   In any case, the best-fit empirical spectrum is the low-gravity M5$\gamma$ 2MASS J0259-4232 object in the 20--40\,Myr-old Columba association \citep{rodriguez2013}. 

\begin{figure*}
\centering
\setlength{\unitlength}{\textwidth}
\begin{picture}(1,0.5)
  \put(0.00,0.00){\includegraphics[width=0.49\textwidth]{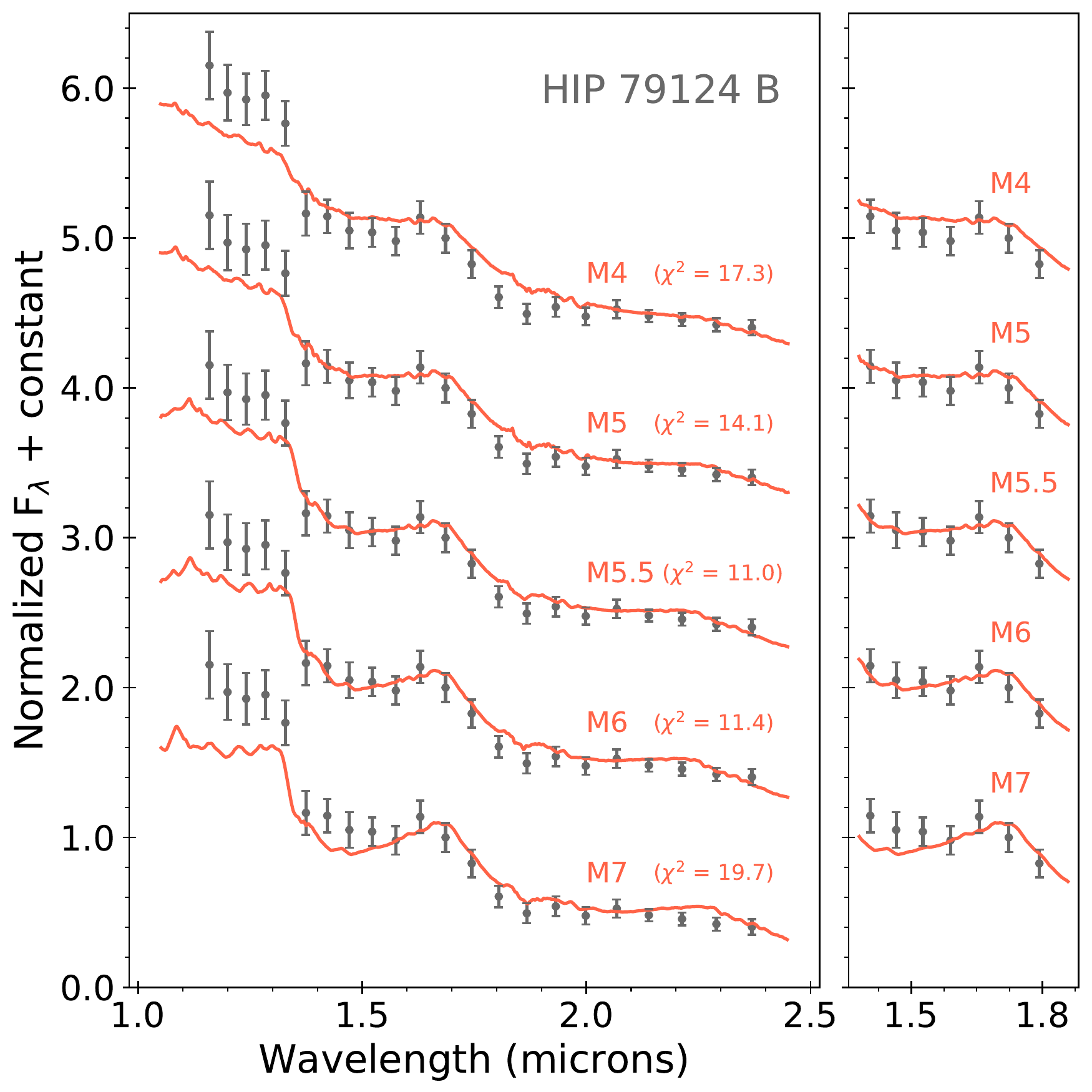}}
 
  \put(0.5,0.00){\includegraphics[width=0.49\textwidth]{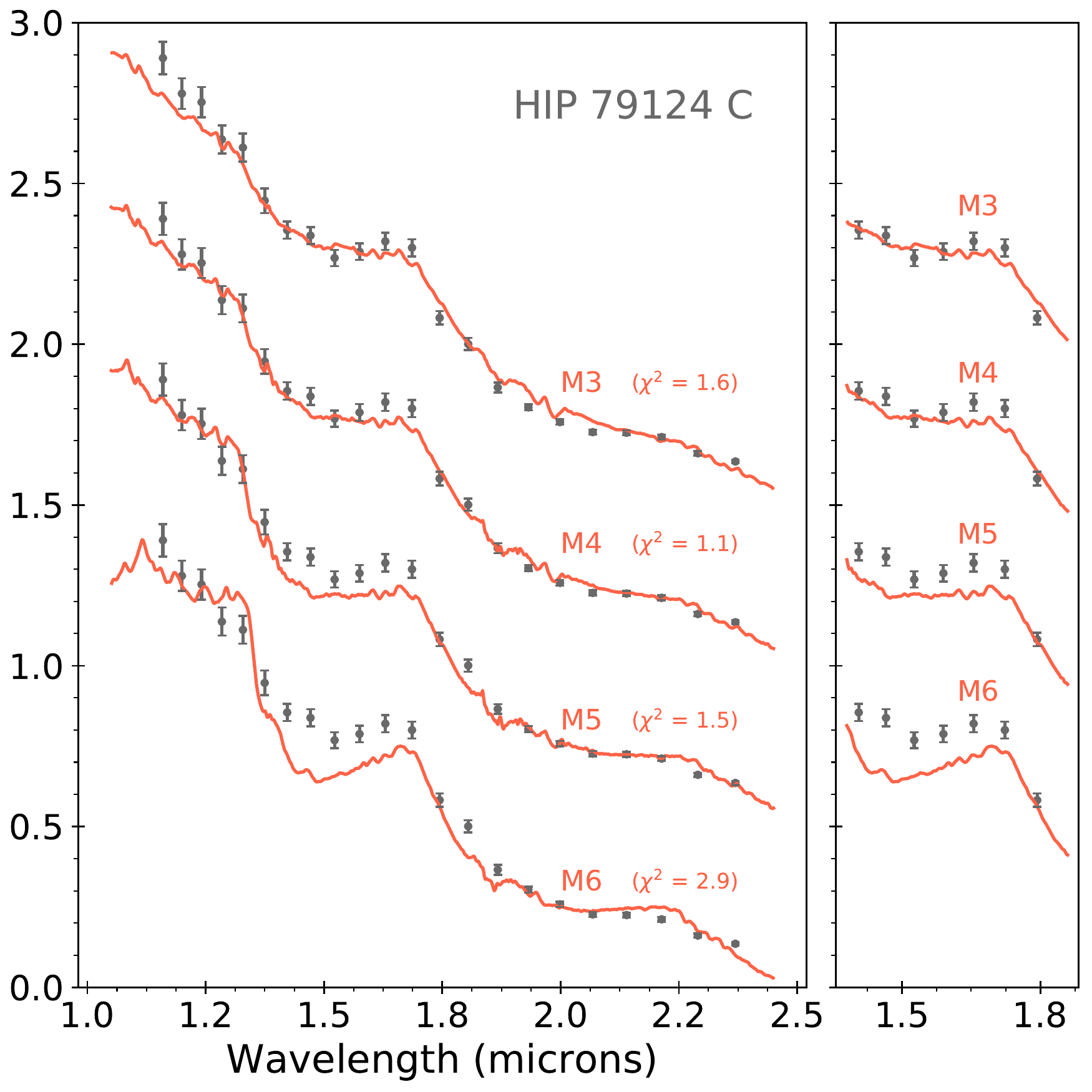}}
 
\end{picture}
\caption{Spectral fits of the B (left) and C (right) companions to the old population ($\sim$\,10\,Myr) of standard spectra from \citet{luhman2017}. Our companions have been corrected for reddening assuming the same extinction as the primary star (see Section \ref{sec:specandphot}). To compute the $\chi^{2}$ per degree of freedom, we refrain from using spectral regions affected by tellurics. The panels on the right of each B and C comparison is an $H$-band zoom-in, which is more affected by the gravity of the object.}
\label{fig:standard}
\end{figure*}

The situation for the inner companion is however more complicated, as the large correlation among the channels yield $\chi^{2}$ results that are higher than in the non-correlated scenario. It also broadens the $\Delta$\,$\chi^{2}$ space of good-fit models (see \cite{greco2016}), which we clearly see in the bow-shape distribution for HIP 79124 B. In this case none of the empirical spectra fall within the 95\,$\%$ confidence region. For this reason, we adopt a confidence interval of \( \Delta\,\chi^{2}\,<\,\sqrt{2\,\chi^{2}\rm_{min}} \) \citep[e.g.,][]{thalmann2013}, which encompasses spectral types in the range M5--M7.


\subsubsection{Comparison to M-type composite standard spectra}
\label{sec:standard}

Here, we adopt the dereddened near-IR standard spectra constructed by \citet{luhman2017}, where they combine several optical spectra for each subtype in the M spectral region. These resulting templates are representative of young associations and can be used for classifying the spectral type of young stars. We take the set of templates produced from a population of objects members of both USco and the TW Hya association (TWA). TWA is located at $\sim$\,50\,pc and, like USco, it has an estimated age of $\sim$\,10\,Myr \citep{webb1999, mamajek2005, donaldson2016}. Following the same procedure as for the library of empirical spectra, we compare the spectrum of HIP 79124 B and C with this set of M-type standard spectra. The results are shown in Figure \ref{fig:standard}. The outer companion is well reproduced in the $JHK$ bands by the M3--M5 spectral standards. The best-fit falls in the M4 type, which is particularly successful at duplicating the $H$-band part of the spectrum. The B companion finds a clear minimum in the M5.5--M6 spectral type regime, where both the $HK$ are very well matched.\par

\begin{figure}
\setlength{\unitlength}{\textwidth}
\hspace*{-0.cm}                                                                                                  
 \includegraphics[width=0.47\textwidth]{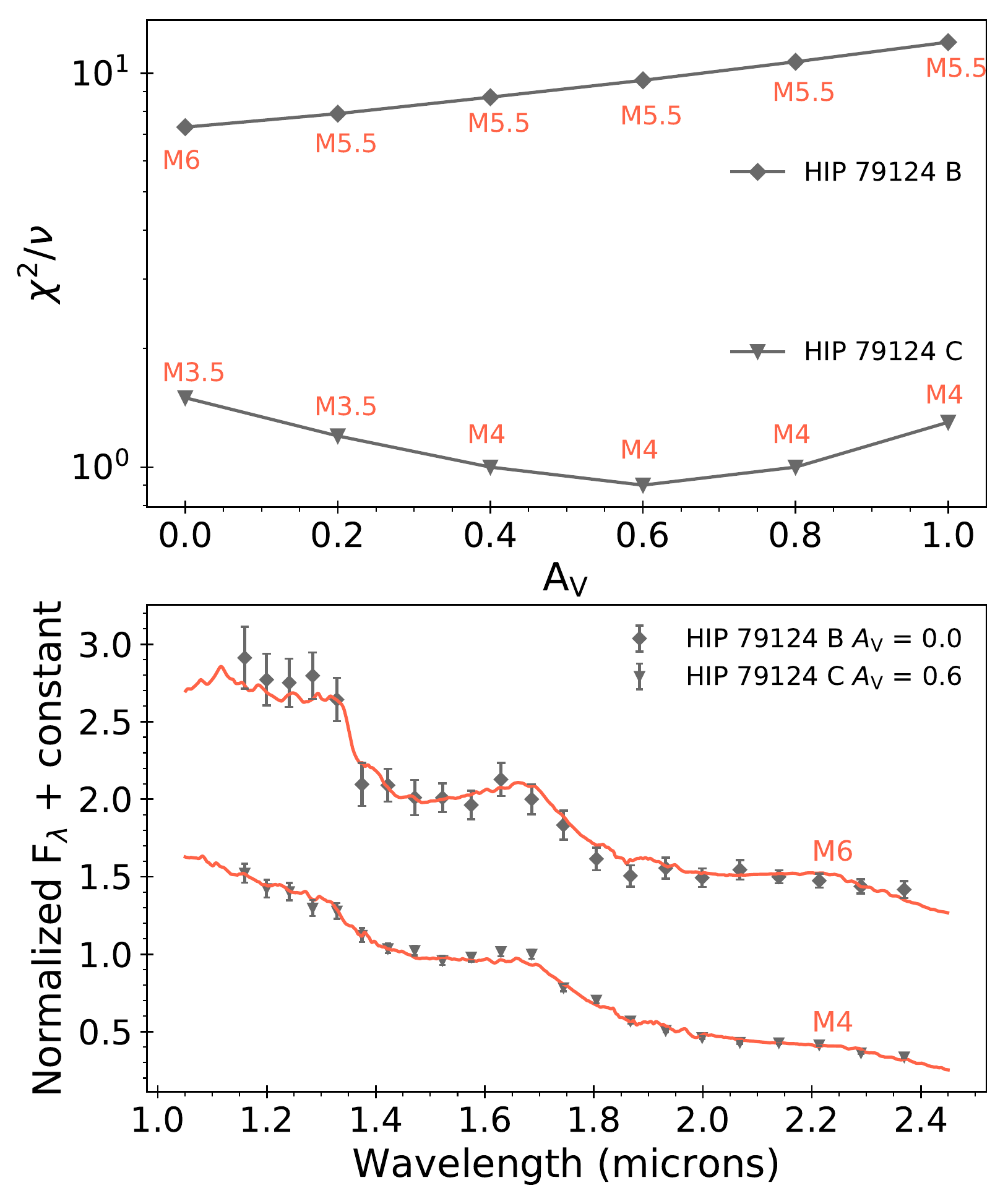}

\caption{(Up) Normalized best-fit $\chi^{2}$ for the B and C SCExAO/CHARIS spectra for a range of different extinctions. The spectral type that corresponds to the best-fit for each individual extinction value is indicated next to each data point. (Down) SCExAO/CHARIS B and C spectra compared to the best-fit standard from \citet{luhman2017}. These SCExAO/CHARIS spectra have been corrected for the extinction value that better minimized the residuals as found in the top panel.}
\label{fig:spectra_reddening}
\end{figure}

The $J$ passband of the CHARIS spectrum of B is slightly brighter than the standard spectra, and it is so consistently for all the  spectral types fitting well the $HK$ bands. This might be due to speckle contamination at the shortest wavelengths by the primary star, given the small projected distance at which HIP 79124 B is located. The Strehl ratio (for a given residual wavefront error) and S/N are also lower at these wavelengths, which ultimately may lead to a suboptimal spectral extraction. Another possibility is that we overestimated the extinction of HIP 79124 B, as in some cases components of young multiple systems have different reddening factors (Kevin Luhman, priv. comm.). We thus adopt a series of different A$\rm_{V}$ and compute their corresponding A$\rm_{J}$, A$\rm_{H}$ and A$\rm_{K_{s}}$ extinction factors using the relations from \citet{fiorucci2003}. In the upper panel of Figure \ref{fig:spectra_reddening} we show the best-fit standard spectrum for each A$\rm_{V}$ value for both HIP 79124 B and C. In the lower panel we show the best-fit spectrum for the combined extinction-spectral type set of parameters. The extinction that minimizes the residuals for the outer companion seems to agree quite well with the derived reddening factor of the primary star (A$\rm_{V}$ = 0.82). However, a extinction-free scenario for B reproduces the SCExAO/CHARIS data much better. As the spectral types are consistent with those obtained in Figure \ref{fig:standard}, for the rest of the calculations we will assume that the computed reddening for the primary star applies to the triple system. \par

\begin{figure}
\centering
\setlength{\unitlength}{\textwidth}
\centering
\begin{picture}(1,0.65)
  \put(0,0.328){\includegraphics[width=0.44\textwidth]{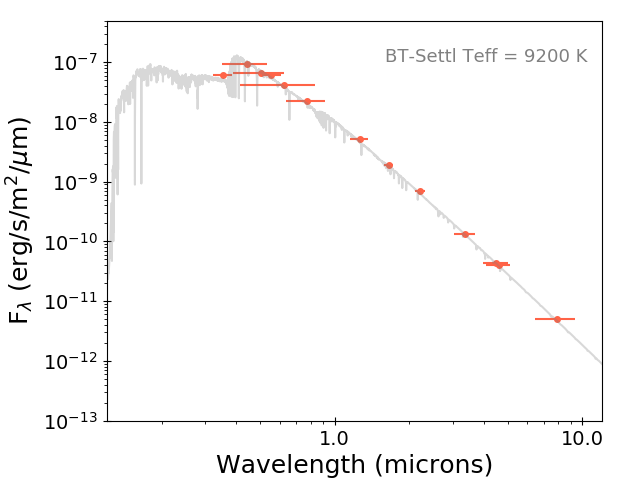}}
  \put(0.016,0.){\includegraphics[width=0.44\textwidth]{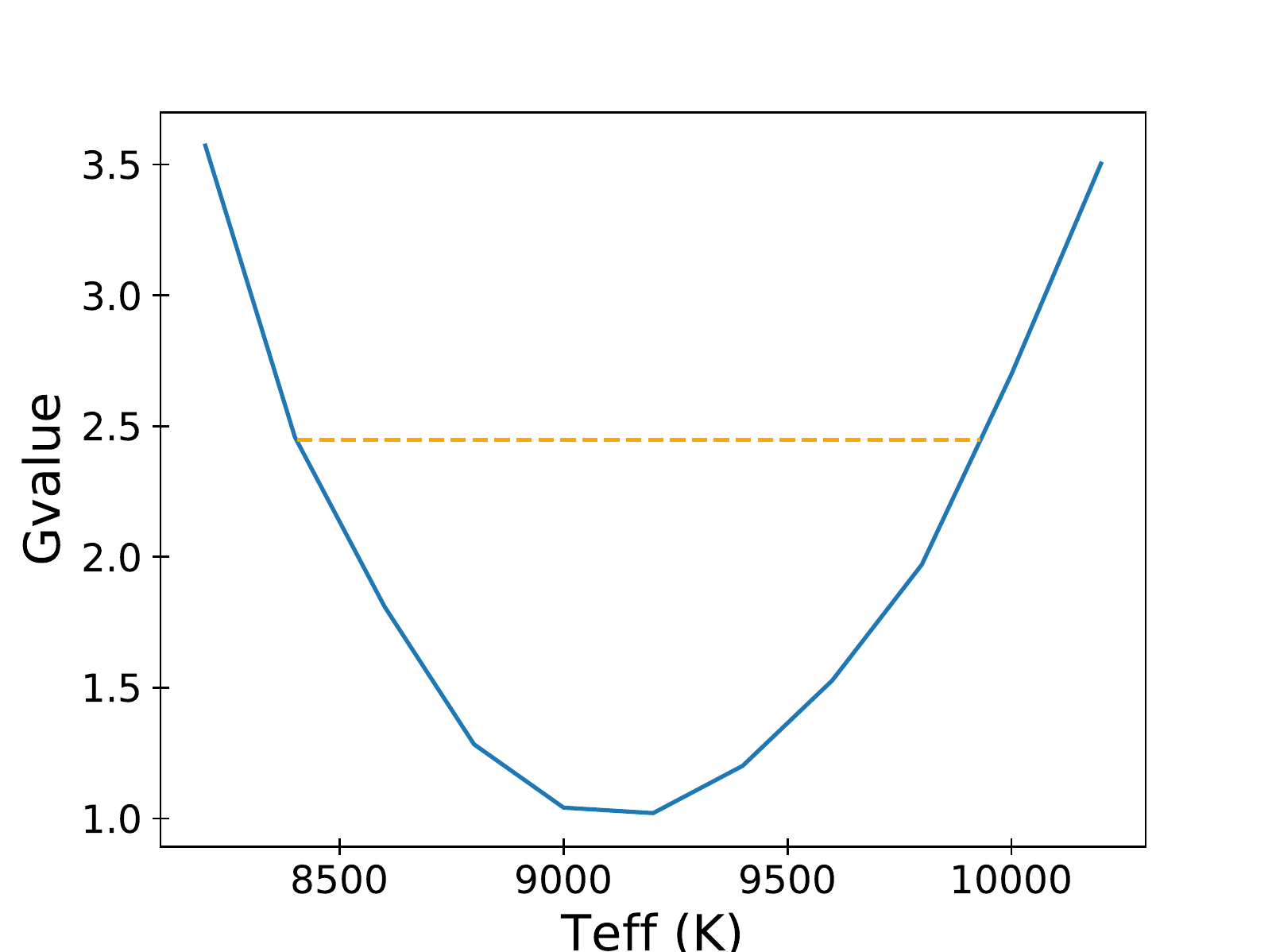}}
\end{picture}
\caption{Spectral energy distribution (SED) of the primary HIP 79124 A0V star. (Top) A BT-Settl model \citep{allard2012} of Teff = 9200\,K and \logg\, = 4.5\,dex is fitted to the flux values at $\leq$\,10\,$\mu$m compiled from the literature (see Table \ref{tab:hip79124ased}). (Bottom) G-value \citep{cushing2008} for BT-Settl models with different T$\rm_{eff}$. The orange dashed line shows the upper limit for the considered well-fitting models.  }
\label{fig:primary_SED}
\end{figure}

\subsubsection{Final adopted spectral types}

Based on the comparison with empirical spectra from the Montreal library and with the \citet{luhman2017} standard templates, here we summarize the final spectral types and uncertainties that we  adopt for the companions. For the C component, from Section \ref{sec:empirical} we obtained a 95\,$\%$ confidence level for M3--M5.5 spectral types, with a dearth of young objects in the Montreal library for spectral types earlier than M5, and an M4 best-fit from Section \ref{sec:standard}. We adopt a final spectral type of M4\,$\pm$\,0.5, which accurately reproduces the $JHK$ passbands of $\sim$10\,Myr standard spectra (see Figure \ref{fig:standard}). For the inner B companion we adopt a spectral type of M6\,$\pm$\,0.5. The Montreal objects favor an M6-type, and a minimum at M5.5--M6 is found using the standard templates. \par

\subsection{HR diagram: age of the HIP 79124 triple system }

Once we have an estimation of the spectral type for each object in the triple system, we place its members on a Hertzsprung-Russell (HR) diagram to compare their position with theoretical models. This will hopefully allow us to constrain the age of the coeval system using such a diverse range of masses.\par

We estimate effective temperatures (T$\rm_{eff}$) and bolometric corrections (BC) from two different sources in the literature of young objects; \citet{kraus2007} presented a set of spectral type models optimized with empirical data from the open cluster Praesepe, with an age of 600\,Myr. More recently, \citet{pecaut2013} used young moving group (5--30\,Myr) members to also build a T$\rm_{eff}$ scale by comparing their spectral energy distribution to BT-Settl atmospheric models \citep{allard2012}, covering spectral types down to M5.\par


\subsubsection{Temperature and luminosity of the companions}
The estimated M4\,$\pm$\,0.5 spectral type for the C companion corresponds to a T$\rm_{eff}$ = 3160\,$\pm$\,140\,K and  BC$_{J}$ = 1.91 $\pm$ 0.05\,mag from the table of young objects in \citet{pecaut2013}. From the SCExAO/CHARIS dereddened absolute magnitude in $J$ band shown in Table \ref{tab:photometryC}, using the BC$\rm_{J}$ value we get a M$\rm_{BOL}$(C) = 7.27 $\pm$ 0.07\,mag, which translates into a luminosity of log($L$(C)/\(L_\odot\)) = -1.01 $\pm$ 0.03\,dex for a solar absolute magnitude of 4.755\,mag.\par

Similarly for HIP 79124 B, we obtain a T$\rm_{eff}$ = 2840\,$\pm$\,90\,K  from \citet{kraus2007}. \citet{pecaut2013} do not count with values for young stars beyond M5, but the error bar encompasses their young M5 T$\rm_{eff}$, and a spectral type later than M6.5 is hardly a good fit to the data, as seen in Section \ref{sec:standard}. Given that no bolometric color correction is available for such a late spectral type in  \citet{pecaut2013}, and that there is the possibility of some contamination from the primary star at short wavelengths, we obtain a BC$_{K}$ = 
3.03\,$\pm$\,0.13\,mag from \citet{golimowski2004} for the best-fit field M6 dwarf. That leads to M$\rm_{BOL}$(B) = 9.2 $\pm$ 0.2\,mag and log($L$(B)/\(L_\odot\)) = -1.77 $\pm$ 0.08\,dex.
\par

\begin{figure*}
\centering
\setlength{\unitlength}{\textwidth}
\begin{picture}(1,0.7)
  \put(0.16,0.00){\includegraphics[width=0.32\textwidth]{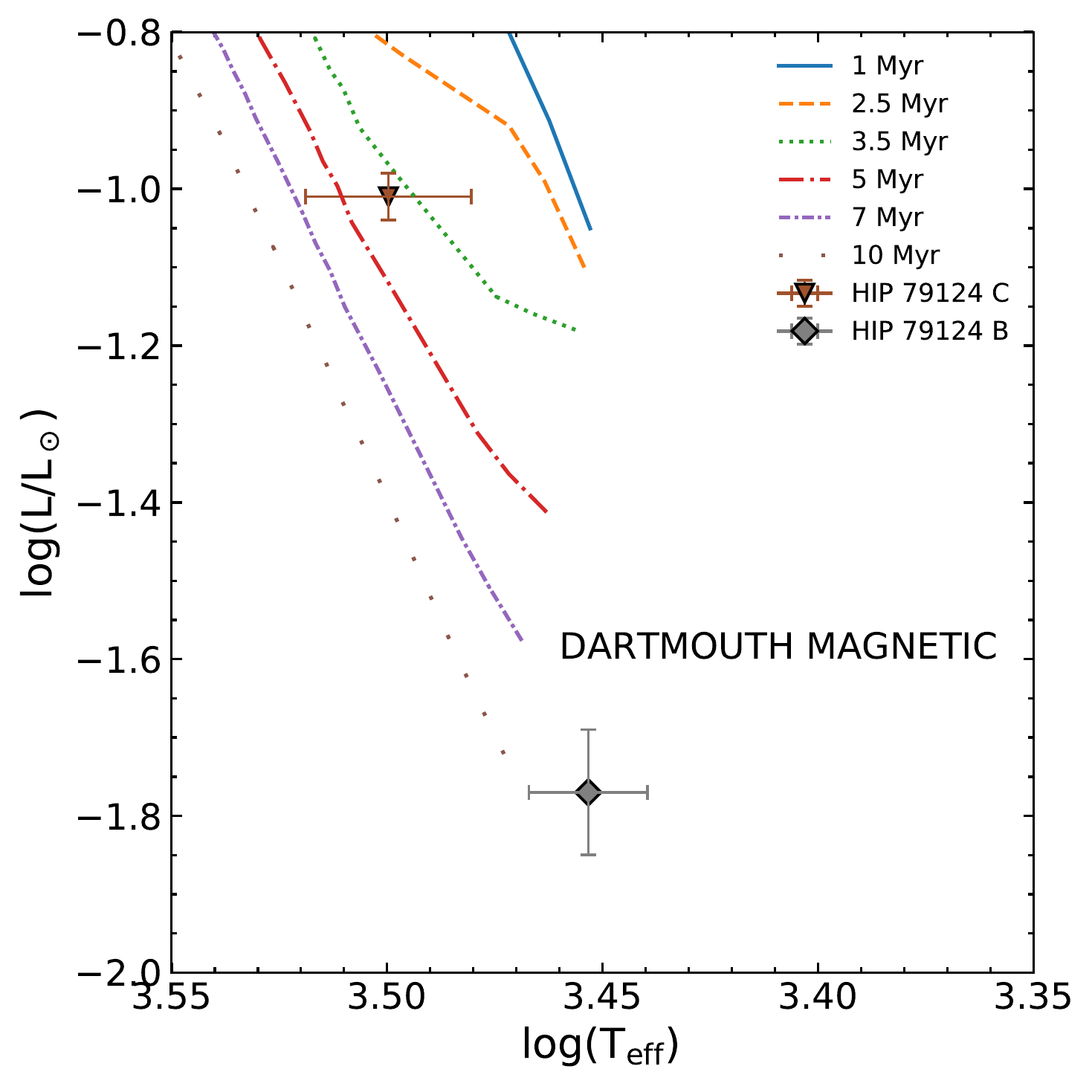}}
 
  \put(0.66,0.35){\includegraphics[width=0.32\textwidth]{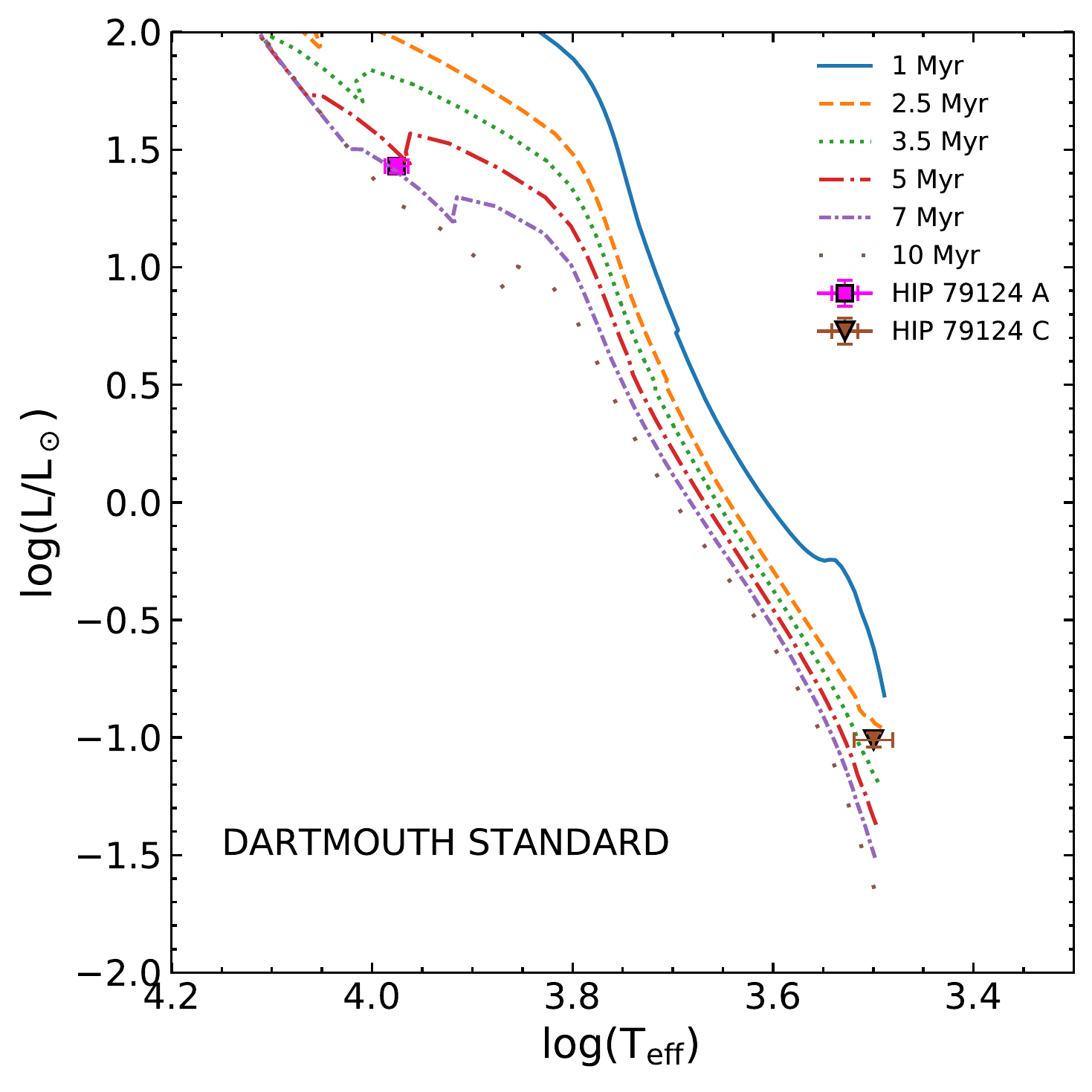}}
  
   \put(-0.0,0.35){\includegraphics[width=0.32\textwidth]{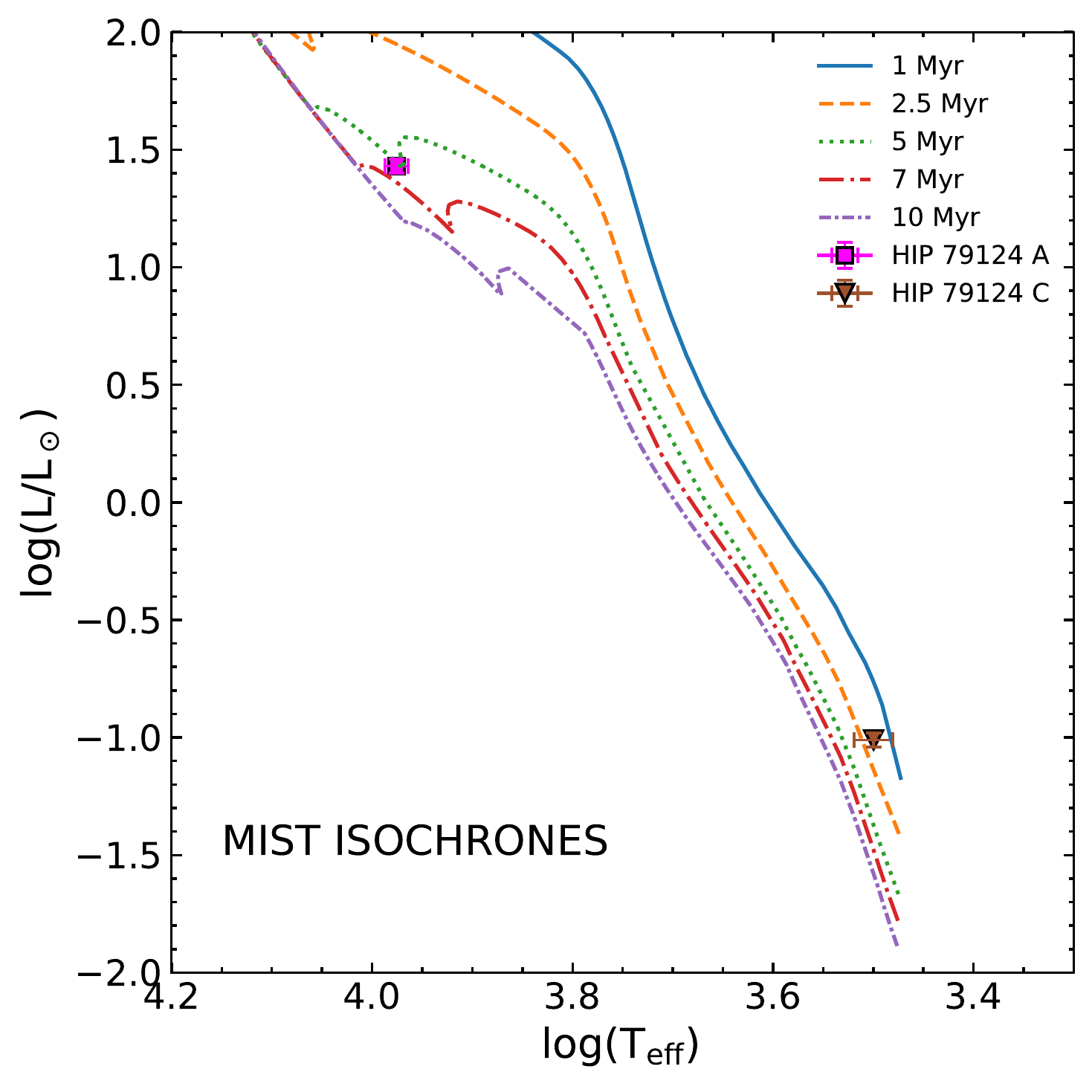}}

   \put(0.33,0.35){\includegraphics[width=0.32\textwidth]{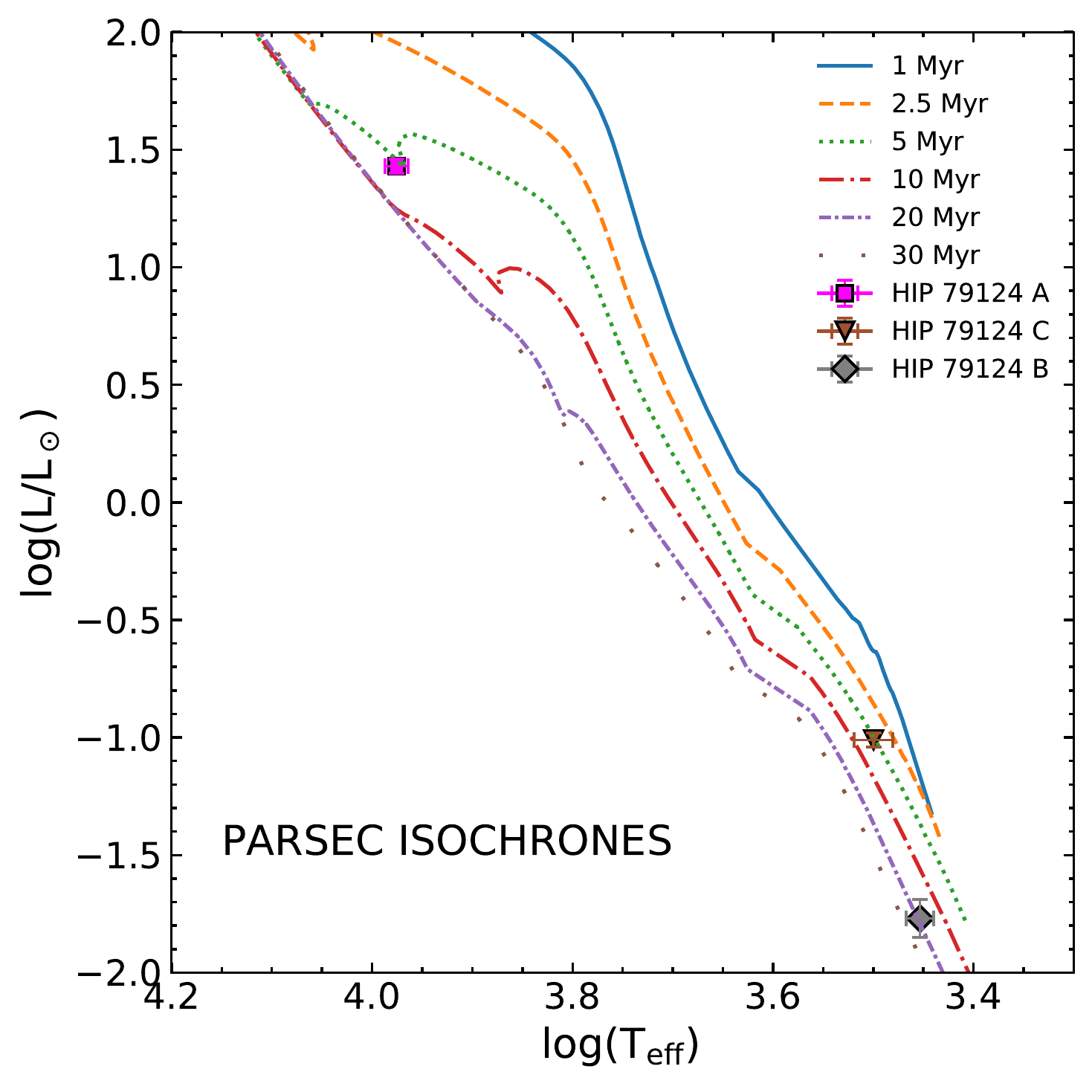}}

   \put(0.55,0.0){\includegraphics[width=0.32\textwidth]{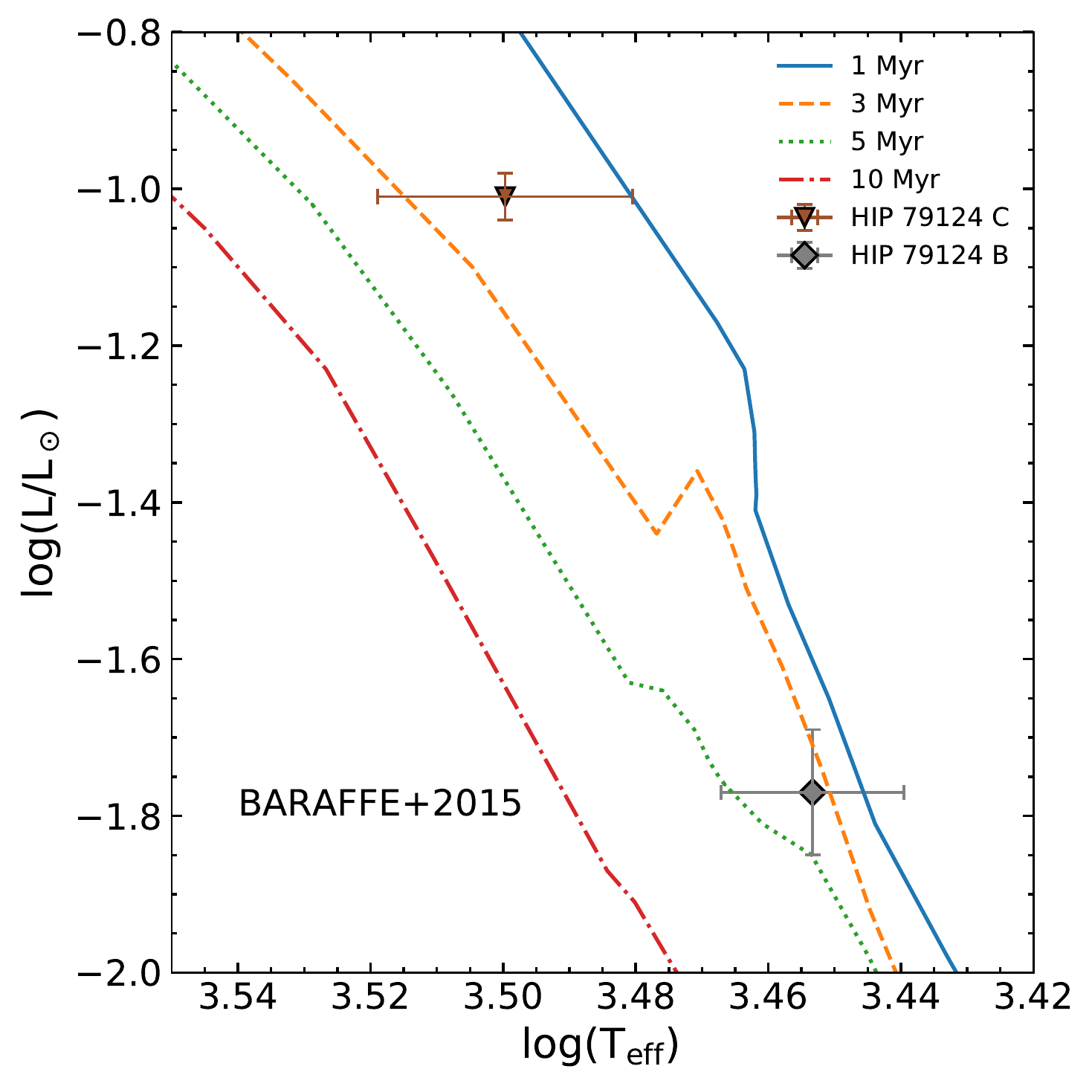}}
 
\end{picture}
\caption{Hertzsprung-Russell diagrams for the HIP 79124 triple system. The observed luminosity and temperature of the individual objects are tested against several theoretical models to derive an age estimate for the system. Numerical values are presented in Table \ref{tab:hip79124age}.  }
\label{fig:age}
\end{figure*}

\begin{table}[!htbp]
\caption{HIP 79124 A SED OBSERVATIONS \label{tab:hip79124ased} }
\centering
\small
\begin{tabular}{lllc }
\hline\hline
Band          & Wavelength      &  Flux &  Reference    \\
 	            	&	\,$\pm$\,Bandwidth 	 &   \scriptsize{(10$^{-10}$\,erg\,s$^{-1}$\,cm$^{-2}$\,$\mu$m$^{-1}$)}   &             \\
\hline                      
U           & 0.3620\,$\pm$\,0.1380 & 242.469\,$\pm$\,11.628 &  (1, 2)    \\ 
B           & 0.4412\,$\pm$\,0.1816 & 399.360\,$\pm$\,19.153 &  (1, 2)    \\ 
G$\rm_{bp}$ & 0.5050\,$\pm$\,0.2347 & 303.500\,$\pm$\,0.376  &  (3, 4)    \\ 
V           & 0.5529\,$\pm$\,0.1129 & 290.463\,$\pm$\,6.965 &  (1, 2)    \\ 
G           & 0.6230\,$\pm$\,0.4183 & 209.466\,$\pm$\,0.077 &  (3, 4)    \\ 
G$\rm_{rp}$ & 0.7730\,$\pm$\,0.2757 & 130.538\,$\pm$\,0.181  &  (3, 4)    \\ 
J           & 1.2603\,$\pm$\,0.2095 & 41.282\,$\pm$\,1.980 &  (1, 5)    \\ 
H           & 1.6652\,$\pm$\,0.1362 & 17.389\,$\pm$\,1.501 &  (1, 5)    \\ 
K           & 2.2094\,$\pm$\,0.2142 & 6.287\,$\pm$\,0.271 &  (1, 5)    \\ 
W1          & 3.350\,$\pm$\,0.660 & 1.347\,$\pm$\,0.045 &  (6) \\
IRAC 2      & 4.4930\,$\pm$\,1.0200 & 0.443\,$\pm$\,0.005 &  (7, 8) \\
W2          & 4.600\,$\pm$\,1.040 & 0.398\,$\pm$\,0.007 &  (6) \\
IRAC 4      & 7.8720\,$\pm$\,2.8810 & 0.0499\,$\pm$\,0.0004 &  (7, 8) \\

\hline
\end{tabular}
\tablebib{(1) \citealt{mann2015}; (2) \citealt{myers2015}; (3) \citealt{jordi2010};
(4) \citealt{gaia2018}; (5) \citealt{cutri2003}; (6) \citealt{cutri2012}; (7) \citealt{quijada2004}; (8) \citealt{carpenter2006}}

\end{table}

\subsubsection{Temperature and luminosity of the host star}

In this case, we take two different approaches. First, as done for the B and C companions, from \citet{pecaut2013} we can obtain a T$\rm_{eff}$ = 9700\,$\pm$\,700\,K with an uncertainty of one subclass in the spectral type. 
As we have a well defined reddening factor in the $V$ band, we use it to calculate a bolometric magnitude M$\rm_{BOL}$(A) = 1.1\,$\pm$\,0.2. We adopted the $V$ apparent magnitude from \citet{hink2015}, the GAIA-DR2 distance \citep{lindegren2018} and BC$_{V}$ for A0 stars in \citet{pecaut2013}. This value is compatible within error bars with the result presented in \citet{hink2015}, although our computation is slightly brighter, as we have adopted the updated GAIA-DR2 distance. This magnitude corresponds to log($L$(A)/\(L_\odot\)) = 1.46\,$\pm$\,0.08\,dex.\par

On the other hand, we attempt to refine these values by constructing the spectral energy distribution of HIP 79124 A from the data in the literature (see Table \ref{tab:hip79124ased}). 
We deredden the observed photometry by using a second-order polynomial fit to the values derived in Section \ref{sec:specandphot},
and refrain from using photometric points at wavelengths $\geq$\,10\,$\mu$m, as they might be affected by the two low-mass companions. 
We then fit BT-Settl models \citep{allard2012} with \logg\, = 4.5\,dex and $[M/H] = 0$, with solar reference abundances from \citet{caffau2011}, to the data via the G goodness-of-fit statistic presented in \citet{cushing2008}, which accounts for the individual filter widths. 
As the reddening factor for the entire wavelength range is uncertain (especially for the shortest wavelengths), we simply adopt an error bar of 5\,$\%$ the flux of each passband. 
The results are presented in Figure \ref{fig:primary_SED}. There is a minimum at 9200\,K. Well-fitting models are again taken if their G values are smaller than $G\rm_{min} + \sqrt{2G\rm_{min}}$. That signifies an uncertainty of 600\,K, probably due to the absence of photometric values with $\lambda$\,$<$\,0.36\,$\mu$m and the effect of reddening.
On the other hand, integrating the best-fit BT-Settl spectrum of T$\rm_{eff}$ = 9200\,$\pm$\,600\,K and using the GAIA-DR2 distance, we obtain log($L$(A)/\(L_\odot\)) = 1.40\,$\pm$\,0.07\,dex \par

The two different approaches are clearly consistent with each other within error bars. To place HIP 79124 A in a HR diagram, we adopt the mean of the resulting values, and their scatter as uncertainty. In this way, we consider for the primary a T$\rm_{eff}$ = 9450\,$\pm$\,250\,K and log($L$(A)/\(L_\odot\)) = 1.43\,$\pm$\,0.03\,dex for HIP 79124 A.

   \begin{figure}
   \centering
   \includegraphics[width=\hsize]{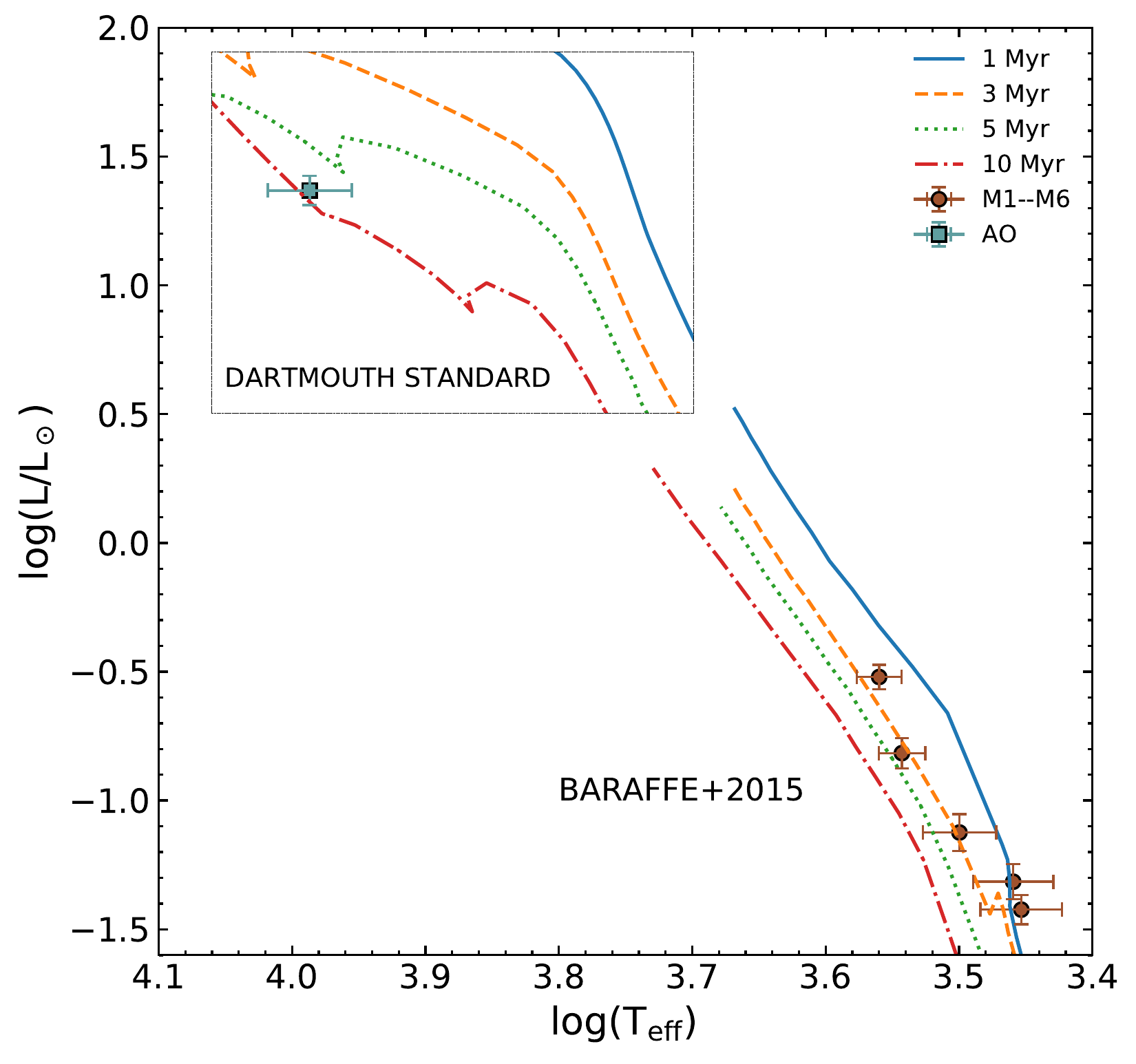}
      \caption{Hertzsprung-Russell diagram for the five M-type stars found within 10\,$\arcmin$ from the location of HIP 79124 A. The BHAC15 isochrones \citep{baraffe2015} are used to estimate their age. We also include the closest A0-type star and confront it against the Dartmouth standard models \citep{dotter2008}. The inner and the outer figures share axes.
              }
         \label{fig:age_vicinity}
   \end{figure}

\subsubsection{HR diagrams}

The HIP 79124 system includes one high-mass star, just arrived (or about to) in the main sequence, and two low-mass pre-main sequence companions contracting along a Hayashi track \citep{siess2000,pecaut2012}. 
The different physical processes occurring on these objects might pose difficulties to the pre-main sequence evolution models, which (if accurate) should be expected to yield a common age estimate for the entire system.    For the similar, albeit older, HD 1160 triple system studied by \citet{garcia2017}, the primary A-type star was on/just beginning to evolve off the main sequence and its two low-mass (M star) companions were closer to the main sequence.   In this case, isochrone comparisons for the primary yielded younger age estimates than those for the two low-mass companions.    Our study allows a similar analysis at young ages.
\par 

Figure \ref{fig:age} shows the luminosity-T$\rm_{eff}$ diagram of the HIP 79124 triple system, where the objects are compared with evolutionary models at different ages. We use a different set of solar-metallicity evolutionary tracks to derive a range of well-fitting ages: the MESA Isochrones $\&$ Stellar Tracks \citep[MIST,][]{paxton2011,paxton2013,paxton2015, choi2016, dotter2016}
for the massive primary star and HIP 79124 C; the PARSEC-COLIBRI stellar isochrones \citep{marigo2017} encompassing the triple system; both the original Dartmouth
Stellar Evolution isochrones \citep{dotter2008} and those accounting for magnetic inhibition of convection\footnote{\url{http://github.com/gfeiden/MagneticUpperSco/}}, which should be reliable for systems that are approximately 10\,Myr, as the surface magnetic field strengths in those models were tuned for modeling $\sim$\,10 \,Myr systems \citep{feiden2016}; and the BHAC15 \citet{baraffe2015} evolutionary tracks for low-mass objects for the B and C companions.   

The isochrones are sufficiently well-spaced to derive precise age estimates for each component.    Early-type stars evolve onto the main sequence along a Henyey track both horizontally and vertically in an HR diagram. Close to the main-sequence ``turn-on", small errors in the temperature/luminosity can translate into larger uncertainties in the component age.   Fortunately, as shown in Figure \ref{fig:age}, HIP 79124 A resides in a region on the HR diagram sufficiently away from the MS turn-on for early A stars: differences in predicted luminosities/temperatures for isochrones at 1--10\,Myr are significantly larger than measurement uncertainties.    While uncertainties are larger for HIP 79124 BC, the vertical spacing for isochrones for low-mass stars is larger, $\gtrsim$ 0.1-0.2 dex. \par

\citet{pecaut2016} reported an age spread of $\pm$\,7\,Myr for their adopted 10-\,Myr US subregion, finding an age gradient within the subgroup where stars are older as they blend with UCL. This seems to be consistent with a star-formation history that might explain the conflict in the derived ages between hot and cold stars \citep{fang2017}. If this was indeed the case, the models would deliver the same age estimates for the three HIP 79124 objects. However, as presented in Table \ref{tab:hip79124age}, we find that the primary seems to be consistently older for the models ($\sim$\,6\,Myr) than the B and C components ($\sim$\,3\,Myr). The PARSEC-COLIBRI isochrones differ significantly from the rest of models for the two low-mass companions, as it has also been the case in previous studies \citep[e.g.,][]{kraus2015}. This might be due to the artificial shift to the models made to fit the observed mass-radius relation for low-mass stars \citep{chen2014}. For this reason, we do not consider their derived values to compute the mean age of HIP 79124 B and C.\par

The fact that the models predict a younger age for low-mass objects compared to the massive primary is in line with the results by \citet{pecaut2016}, where essentially all the objects with a later type than M0 are inferred to have an age below 5\,Myr, moving to younger ages as the stars are cooler. Also, an age of 6\,Myr for the primary star is indeed expected from the location of the HIP 79124 system in the USco subregion. Figure 9 in \citet{pecaut2016} shows a map of the spatial distribution of derived median-ages within the Sco-Cen complex. In this diagram, HIP 79124 falls exactly in the northern part of USco where stars tend to be younger than the mean age of the subregion. \par

The magnetic isochrones from \citet{feiden2016}, which take into account the possible magnetic inhibition of convection in young low-mass stars, seem to provide a more compatible age estimate between the A and C components.
Another effect linked to magnetic fields is the occurrence of spots on the surface of young stars. Spotted stars not only cause inflated radii at all ages, but they also experience a decrease in their luminosity and temperature (the latter especially for low-mass stars), which may produce a dispersion in the HR diagram. As found by \citet{somers2015}, this effect makes PMS stars appear younger and less massive when spots are present. Although  the scatter does not seem to be high enough to explain the global age-mass discrepancy, it might be a contributor to take into account. In this way, we can apply the age correction factors derived in \citet{somers2015} to our 3-Myr HIP 79124 B and C companions. Assuming that 1/6 of the stellar surface is covered by spots, we find that C would have an age of $\sim$\,6\,Myr (for a corresponding mass of $\sim$250\,M$\rm_{Jup}$, see Section \ref{sec:mass}). The B companion is not massive enough to derive a correction factor from the models, but its age would certainly lay beyond 6\,Myr. This is very much in line with the observed age of the primary star. Indeed, the fact that models do not reproduce the effect of magnetic fields, which slow down the contraction of PMS stars and affect their luminosity and temperature, appears to explain our results well.\par

Another source of uncertainty in the PMS ages arises from the physical processes that contribute to the initial position of the star in the HR diagram at $t = 0$, from which the contraction follows. The radius at which the contraction starts varies with early accretion rate, which creates a spread in radii (and luminosities) with which the stars of different masses are born \citep[e.g.,][]{hartmann2003, soderblom2014}. This introduces uncertainties on the contraction ages, especially for intermediate-mass stars \citep{hartmann2016}. In the case of HIP 79124 A, a shift in luminosity of log($L/L\rm_\odot$) $\sim$\,0.5\,dex would be required for it to match the $\sim$\,3\,Myr-old isochrone that better fits the age of the low-mass companions. This means that for the triple system to have common ages, the birthline for intermediate-mass stars would need to be corrected to a lower luminosity level by a factor of $\sim$\,3.

Finally, in a similar fashion to what has been done for the HIP 79124 companions, we have obtained an estimation of the luminosity and temperature of the stars in the vicinity (within 10\,$\arcmin$) of HIP 79124 A. We found five M-type stars members of USco confirmed by the BANYAN $\Sigma$ tool \citep{gagne2018}, which are confronted in Figure \ref{fig:age_vicinity} against the BHAC15 isochrones. When no extinction factors were available or we could not calculate them, we adopted the same extinction as for HIP 79124 A, and an uncertainty of 0.5\,mag. If the calculated extinction was non-physical, i.e., negative, A$\rm_{V}$ was set to zero. We have also included the closest massive star to HIP 79124 A, which also happens to be an A0-type star at about 28\,$\arcmin$, compared to the Dartmouth standard models. In the same way as for the HIP 79124 system, the low-mass stars tend to give an age of $\sim$\,3\,Myr, about half of the age that the Dartmouth models estimate for the A0 star (5--10\,Myr). \par

\begin{figure*}
\centering
\setlength{\unitlength}{\textwidth}
\begin{picture}(1,0.8)
  \put(0.1,0.4){\includegraphics[width=0.4\textwidth]{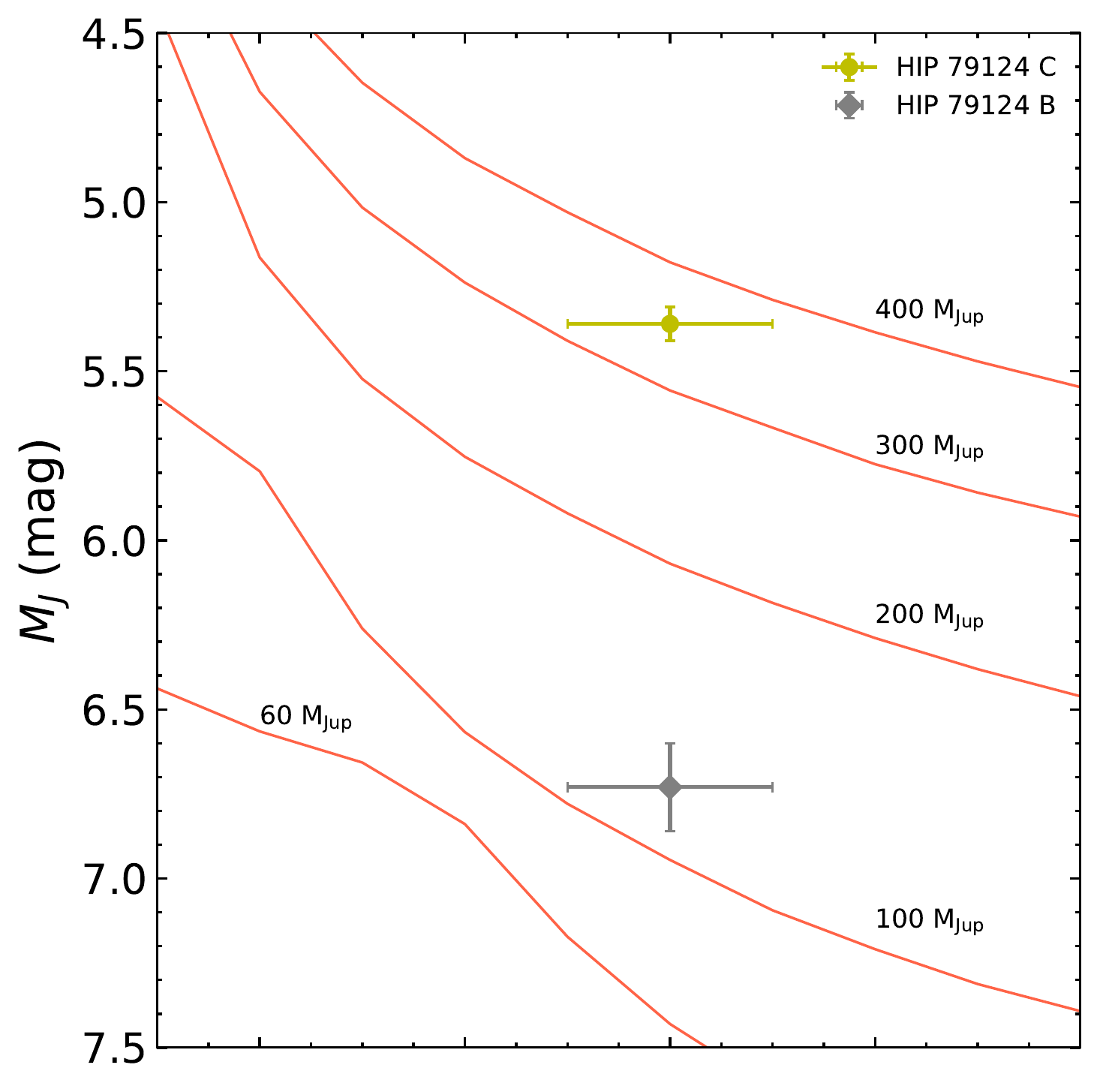}}
  \put(0.5,0.4){\includegraphics[width=0.4\textwidth]{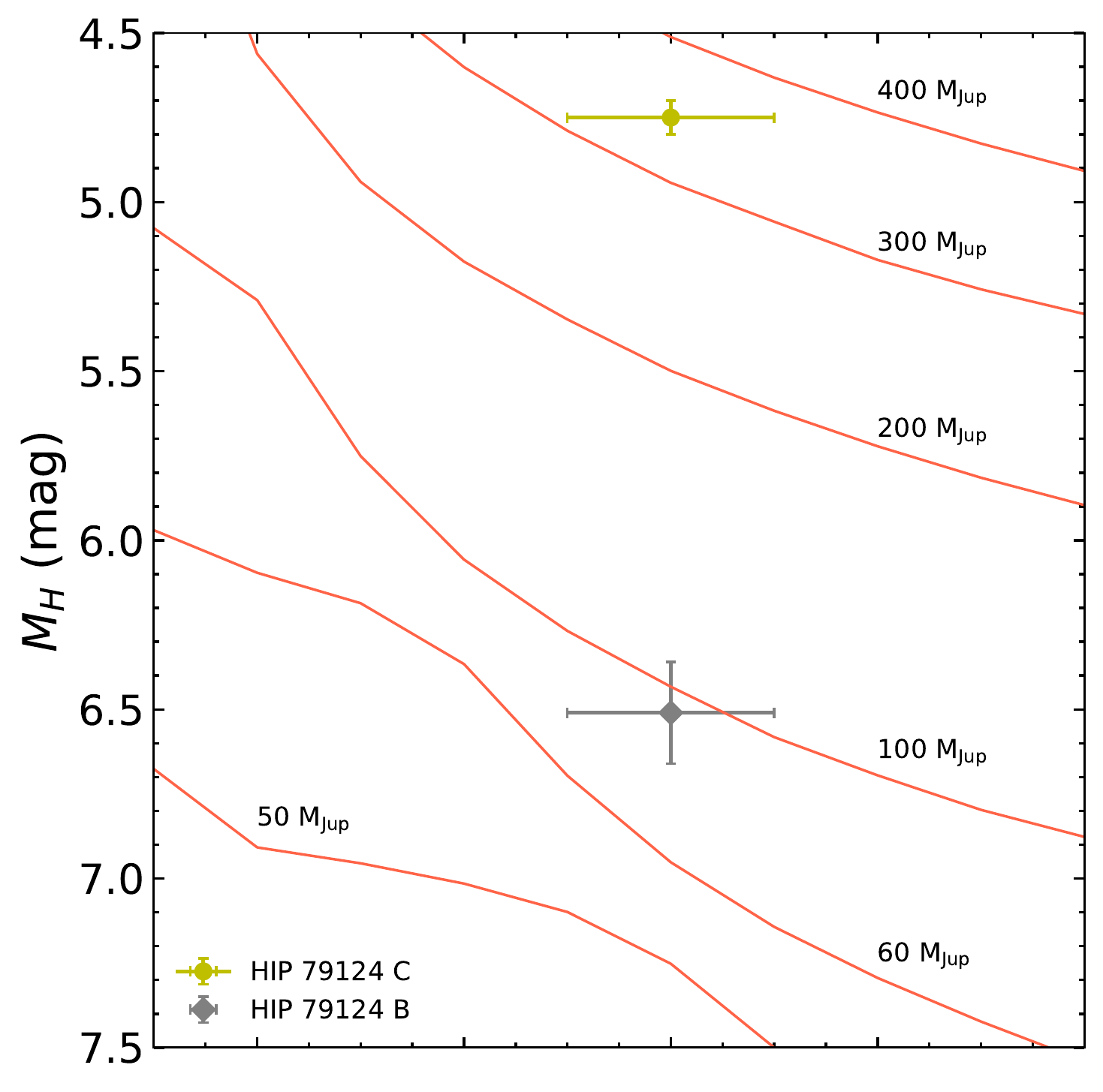}}
 \put(0.1015,0.){\includegraphics[width=0.41\textwidth]{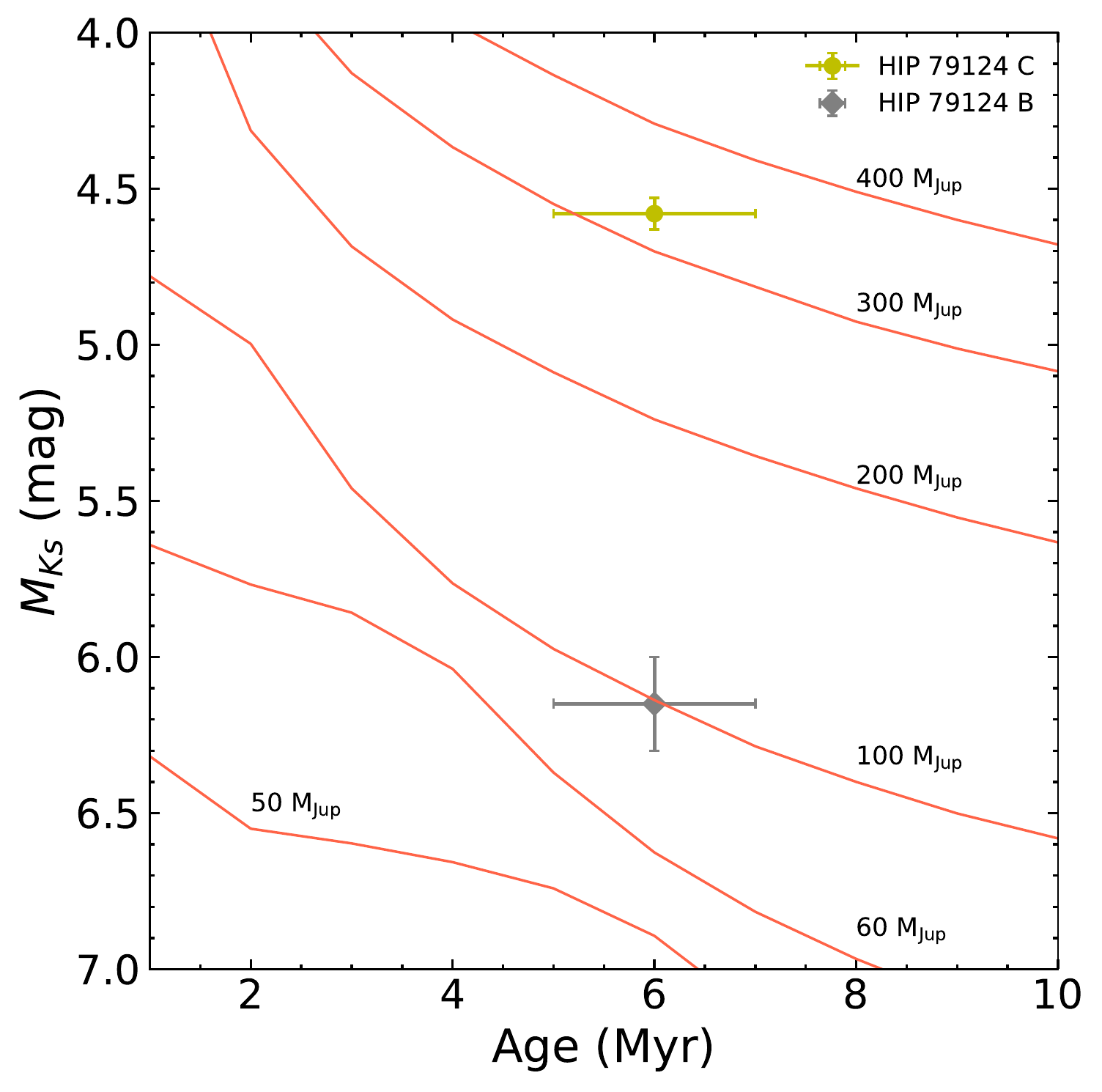}}
  \put(0.499,0.){\includegraphics[width=0.41\textwidth]{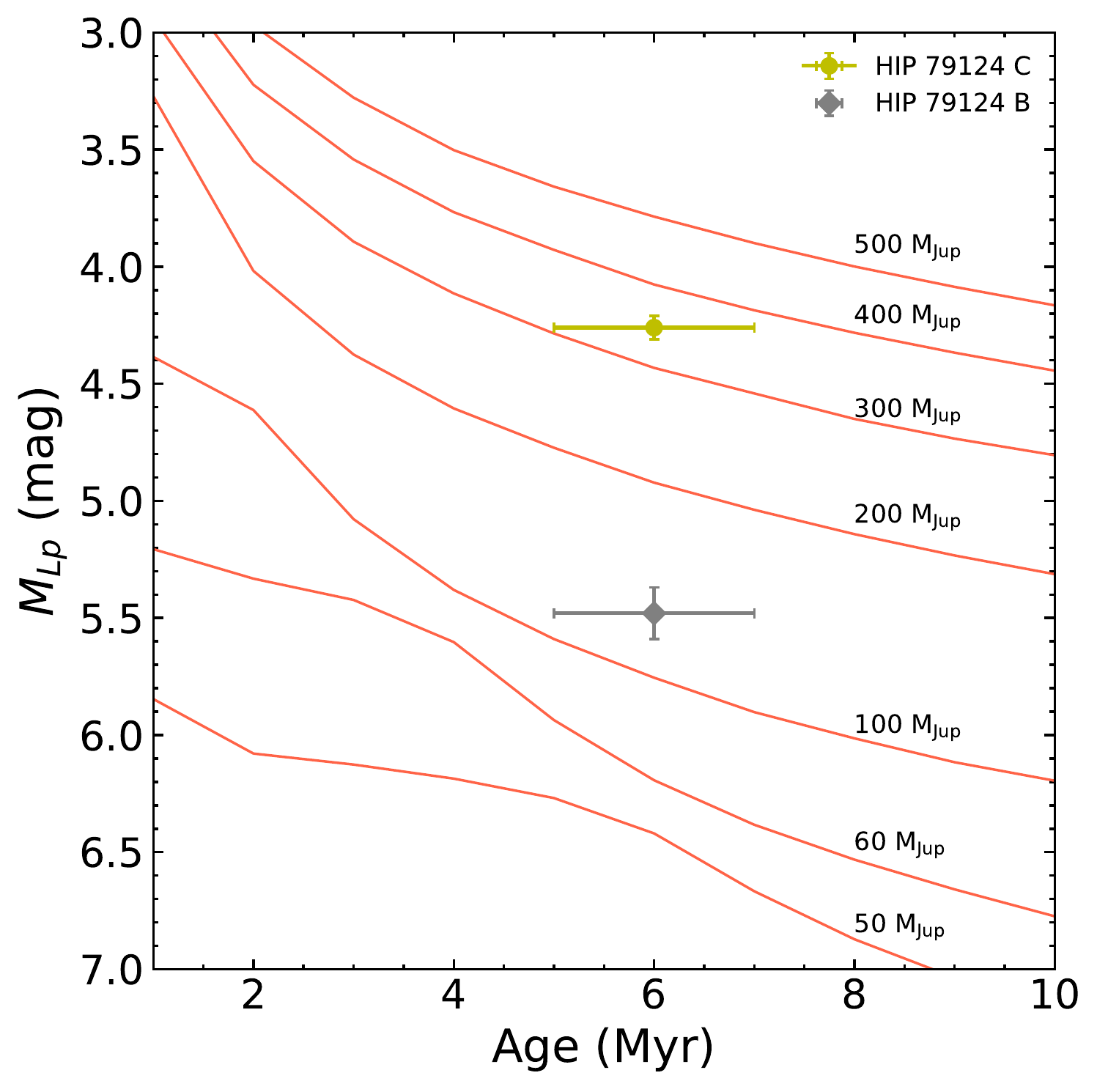}}

\end{picture}
\caption{Absolute magnitude vs Age of the HIP 79124 low-mass comapnions for $JHK\rm_{s}$\Lp. \citet{baraffe2015} models are overplotted to estimate their masses. The age of the companions is set at 6\,$\pm$\,1\,Myr, taken from the primary age estimation (see Fig. \ref{fig:age} and Table \ref{tab:hip79124age}). }
\label{fig:mass}
\end{figure*}

\subsection{Mass of the HIP 79124 triple system}
\label{sec:mass}

We thus consider an age of 6\,$\pm$\,1\,Myr for HIP 79124, derived from the primary A0-type star. With this parameter well constrained, the observed $JHK\rm_{s}L\rm_{p}$ photometry presented in Table \ref{tab:photometryB} and \ref{tab:photometryC} and the excellent accuracy in the distance to the system taken from GAIA-DR2, we can derive the mass of B and C using the BHAC15 isochrones \citep{baraffe2015}.\par

Figure \ref{fig:mass} shows an approximate mass for the low-mass companions using $JH$\Ks photometry derived from our CHARIS data and Keck/NIRC2 \Lp photometry.    The close M6-type B companion agrees well with a mass of $\sim$\,100\,M$ \rm_{Jup}$. \citet{hink2015} reported a mass of $\sim$\,135\,M$\rm_{Jup}$ for an age of 10\,Myr from L$\rm_{p}$ observations, as we similarly obtain for 6\,Myr. 
It is interesting to note that this object seems to be slightly brighter in $J$ and L$\rm_{p}$ bands than in $H$ and \Ks. The effect in $J$ can be explained by the object simply being bright in this band, or it might also be due to slight contamination from the primary star (see Section \ref{sec:standard}), while in $L\rm_{p}$ the object appears to be somewhat red compared to what it is predicted by BT-Settl models \citep{allard2012, baraffe2015}. C is more massive and it is found to fall at $\sim$\,350\,M$\rm_{Jup}$ in all bands, which proves the consistency of our derived $JH$\Ks\,photometry coupled with the extracted $L_{p}$ magnitudes from Keck/NIRC2 archival data. If we considered an age of the system of 3\,Myr, as obtained from the models for the low-mass companions, B and C would have respective estimated masses of $\sim$55\,M$\rm_{Jup}$ and $\sim$250\,M$\rm_{Jup}$. Applying the correction for spotted stars from \citet{somers2015} for the C companion, its mass would be of the order of $\sim$310\,M$\rm_{Jup}$, very close to the mass obtained using the age derived from the primary, which indicates that magnetic fields have an important role in the observed discrepancy. \par

This same effect can be translated to the inferred masses of exoplanets discovered by direct imaging in star-forming regions. For instance, the recent planetary-mass companion revealed within the transition disk around the $\sim$\,5\,Myr-old PDS 70 star \citep{keppler2018} is estimated to have a mass of $\sim$\,5\,M$\rm_{Jup}$ from its photometry, using the hot-start COND models \citep{baraffe2003}. As PDS 70 is a low-mass K7-type star in Upper Centaurus-Lupus group (with a mean age of 16\,$\pm$\,2\,Myr \citep{pecaut2016}), one might hypothesize that the real age of the system could be underestimated. If PDS 70 was older by a factor of 2, as we see for HIP 79124, the planetary-mass companion would be more massive, of $\sim$\,7\,M$\rm_{Jup}$, using again the COND models. Similarly, the ROXs 42B T Tauri binary star with an M0-type primary, is a member of the $\rho$ Oph complex, and hosts a directly imaged companion. According to the COND models, this circumbinary object lies in the planetary-mass regime with $\sim$\,10\,M$\rm_{Jup}$ \citep{currie2014b}. If instead of $\sim$\,2.5\,Myr the system age was twice of that, the mass of the planet would increase up to $\sim$\,13\,M$\rm_{Jup}$.

\begin{table}[!htbp]
\caption{Model age estimates of the triple system (in Myr)\label{tab:hip79124age} }
\centering
\begin{tabular}{lccc }
\hline\hline
Model          & A      &  B &  C    \\
\hline                      

MIST           &  5\,$\pm$\,1 & -- &  2\,$\pm$\,1    \\ 
PARSEC-COLIBRI &     6\,$\pm$\,1 &   17\,$\pm$\,5  & 5$^{+5}_{-2.5}$              \\

Dartmouth std &        6\,$\pm$\,1  & --   &   3\,$\pm$\,1 \\

Dartmouth mag &        --         & --   &     4$^{+3}_{-1}$  \\

Baraffe+2015 &        --         &  3\,$\pm$\,2  &   2\,$\pm$\,1  \\

\hline

\textbf{Mean} &   6\,$\pm$\,1    &   3\,$\pm$\,2    &  3\,$\pm$\,1   \\
\hline

\end{tabular}
\tablefoot{The PARSEC-COLIBRI isochrones are not considered in the computation of the mean age of the B and C low-mass companions (see text). }

\end{table}

\subsection{Formation scenario}

Given the stellar nature of the companions, the natural approach  to study the formation path of the low-mass stars would be to consider the fragmentation of the molecular cloud that gave origin to the HIP 79124 system. Indeed, radiation hydrodynamical calculations by \citet{bate2012} show that the initial mass function (IMF) of the formed objects peak at about the masses of HIP 79124 B and C. These simulations also reproduce a formation timescale for a massive A-type star that is very similar (well below 1\,Myr difference) to that of low-mass objects down to the brown dwarf regime. Triple systems however seem to be rare, with a frequency of $\sim$\,4\,$\%$, as also found by observational results \citep[e.g.,][]{daemgen2015}, and a separation distribution culminating at $\sim$\,100\,AU.\par

An intriguing possibility is that the two low-mass stellar companions are formed via gravitational instability (GI). In this scenario, a massive and cold disk may gravitationally collapse and break down in fragments of sizes ranging from planetary-mass companions to low-mass stars on wide orbits \citep[e.g.,][]{boss1998, rafikov2005}. For one of these fragments to form, the cooling timescale needs to be shorter than the orbital period, which also assures coevality of the triple system in the GI scenario \citep{gammie2001}. Recent models are able to generate a synthetic population of GI-formed objects, and dynamically evolve the system before and after disk dispersal \citep{forgan2013, forgan2015, forgan2018}. These predictions have been tested against high-contrast direct imaging data, showing that, if substellar objects at separations $>$\,30\,AU are indeed formed via GI, this formation method is rare \citep{vigan2017}.\par

 For solar-mass stars with protoplanetary disks extending up to 100\,AU, models by \citet{forgan2018} resulted in the formation of companions as massive as $\sim$120\,M$\rm_{Jup}$. \citet{sta2009} also showed that the fragmentation of a 400\,AU disk around a 0.7\,\(M_\odot\) star can give rise to a broad range of companions, 30\,$\%$ of them being low-mass stars (up to $\sim$\,200\,\,M$\rm_{Jup}$). This indeed could be the formation process to explain the location of HIP 79124 B, a $\sim$100\,M$\rm_{Jup}$ star located at a projected separation of only $\sim$\,25\,AU.
 

\begin{table*}[!htbp]
\caption{Parameters of the HIP 79124 triple system\label{tab:triple_param} }
\centering
\small
\begin{tabular}{c c c c c c c c c }
\hline\hline
HIP 79124           & Distance          & Spectral Type            & T$\rm_{eff}$                 & log($L/L\rm_\odot$)  & Age   & Mass & Sep. & PA  \\
 	            	&	    (pc)       	 &               & (K)         &   (dex)  &   (Myr)         &  (M$\rm_{Jup}$)  &(mas)       &  ($\deg$)      \\
\hline                      

A & 137.0\,$\pm$\,1.2\tablefootmark{a} & A0V\tablefootmark{b}   & 9450\,$\pm$\,250 &  1.43\,$\pm$\,0.03 & 6\,$\pm$\,1 & -- & -- & --    \\

B & -- & M6\,$\pm$\,0.5 & 2840\,$\pm$\,90 & -1.77\,$\pm$\,0.08 & 3\,$\pm$\,2 & 100\,$\pm$\,30 & 180\,$\pm$\,5 & 252.9\,$\pm$\,1.6  \\

C & -- & M4\,$\pm$\,0.5   & 3160\,$\pm$\,140 &  -1.01\,$\pm$\,0.03 & 3\,$\pm$\,1 & 330\,$\pm$\,30 & 967\,$\pm$\,6  & 100.39\,$\pm$\,0.03 \\

\hline
\end{tabular}
\tablefoot{\tablefoottext{a}{From the $Gaia$-DR2 \citep{lindegren2018} }
\tablefoottext{b}{From \citet{houk1988}}

}
\end{table*}

\section{Conclusions}

We have presented the first spectrophotometric study of the USco HIP 79124 triple system with SCExAO/CHARIS. Combining low-resolution $JH$\Ks\, spectroscopy with archival \Lp photometry from \citet{hink2015} and \citet{ser2017}, we
estimate the spectral types of the companions, which altogether serves as a diagnostic to derive the age of the system and the masses of the low-mass objects. The key results of our analysis can be summarized as

\begin{itemize}
\renewcommand\labelitemi{--}
  \setlength\itemsep{0.2em}

\item SCExAO/CHARIS detects HIP 79124 B and C in low-resolution mode without the employment of any PSF-subtraction algorithm at a S/N of $\sim$\,9 and $\sim$\,120 and at distances of $\sim$0.18\,$\arcsec$ and $\sim$0.97\,$\arcsec$, respectively.

\item We account for the correlated noise present in IFS data \citep{greco2016}. B falls in a highly-correlated regime, even for well-separated wavelength channels. In the case of the outer C companion, the uncorrelated amplitude is predominant. Using these correlated errors, we find that young ($\sim$\,10\,Myr) standard objects from \citet{luhman2017} best match the spectra of B (M6) and C (M4).  

\item We assemble an HR diagram where we place the triple system, and confront their luminosity-T$\rm_{eff}$ values with several theoretical models to assess a common age estimate. However, the primary star is found to have an age of $\sim$\,6\,Myr, while the models consistently deliver about half this age for the low-mass companions.

\item This age-mass discrepancy for young low-mass stars is in line with the results seen in several young regions such as Sco-Cen \citep{pecaut2016}. As HIP 79124 should be coeval with the three objects forming in timescales <\,1\,Myr \citep{bate2012}, this result strongly points towards the fact that the models do not reproduce well enough the PMS phase of low-mass stars.

\item Adopting the age of the primary star for the entire system, we find a mass of B of $\sim$\,100\,M$\rm_{Jup}$, and $\sim$\,330\,M$\rm_{Jup}$ for C. Given their masses and small orbital separation, there is the possibility that these objects formed via disk instability \citep[e.g.,][]{forgan2018}. 

\item This effect can alter the mass of the directly-imaged companions to low-mass stars, if the age of the system is derived from isochronal fits to the photometric data of the host star.

\end{itemize}

We have demonstrated the SCExAO/CHARIS capabilities by resolving a very packed system and constraining their properties through low-resolution spectroscopy. For this very likely coeval system, models predict an older age for the A0-type primary star than for the low-mass companions. This result might be related to magnetic field effects, which implies that the models do not reproduce with enough accuracy the contraction rate of low-mass PMS stars or the presence of stellar spots , and thus deliver a younger age \citep{somers2015,feiden2016, somers2017}. Further observations of this system could constrain the orbit of the B companion to derive its period and a more reliable mass estimate.

\begin{acknowledgements}
We thank Kevin Luhman, Eric Mamajek, and Mark Pecaut for helpful draft comments.  We wish to emphasize the pivotal cultural role and reverence that the summit of Maunakea has always had within the indigenous Hawaiian community.  We are most fortunate to have the privilege to conduct scientific observations from this mountain.
R.A.-T. and M.J. gratefully acknowledge funding from the Knut and Alice Wallenberg foundation.  S.D. acknowledges support from the “Progetti Premiali” funding scheme of the Italian Ministry of Education, University, and Research. E. A. is supported by MEXT/JSPS KAKENHI grant No. 17K05399. M.T. is supported by MEXT/JSPS KAKENHI grant Nos. 18H05442, 15H02063, and 22000005.M. H. is supported by the Grant-in-Aid for Scientific Research on Innovative Areas (2302, 23103002) under the Ministry of Education, Culture, Sports, Science and Technology (MEXT) of Japan.
This research has benefited from the Montreal Brown Dwarf and Exoplanet Spectral Library, maintained by Jonathan Gagne. This work has made use of data from the European Space Agency (ESA) mission {\it Gaia} (\url{https://www.cosmos.esa.int/gaia}), processed by the {\it Gaia}
Data Processing and Analysis Consortium (DPAC, \url{https://www.cosmos.esa.int/web/gaia/dpac/consortium}). Funding for the DPAC has been provided by national institutions, in particular the institutions participating in the {\it Gaia} Multilateral Agreement.

\end{acknowledgements}


\end{document}